# Flexocoupling-induced phonons and ferrons in van der Waals ferroelectrics

Anna N. Morozovska[1*], Eugene A. Eliseev[2], Oleksiy V. Bereznikov[1], Mykola Ye. Yelisieiev[3], Guo-Dong Zhao[4], Yujie Zhu[5], Venkatraman Gopalan[4†], Long-Qing Chen[4‡], Jia-Mian Hu[5§], and Yulian M. Vysochanskii[6**]

[1] Institute of Physics, National Academy of Sciences of Ukraine, 46, pr. Nauky, 03028 Kyiv, Ukraine

[2] Frantsevich Institute for Problems in Materials Science, National Academy of Sciences of Ukraine, Omeliana Pritsaka str., 3, Kyiv, 03142, Ukraine

[3] Institute of Semiconductor Physics, National Academy of Sciences of Ukraine, 45, pr. Nauky, 03028 Kyiv, Ukraine

[4] Department of Materials Science and Engineering, Pennsylvania State University, University Park, PA 16802, USA

[5] Department of Materials Science and Engineering, University of Wisconsin-Madison, Madison, WI, 53706, USA

[6] Institute of Solid-State Physics and Chemistry, Uzhhorod University, 88000 Uzhhorod, Ukraine

**Abstract**

The contribution of flexoelectric coupling to the long-range order parameter fluctuations in ferroics can be critically important to the ferron dispersion and related polar, pyroelectric and electrocaloric properties. Here we calculate analytically the dispersion relations of soft optic and acoustic flexocoupling-induced phonons and ferrons by incorporating the flexoelectric coupling, damping, and higher elastic gradients in the Landau-Ginzburg-Devonshire free energy functional using the van der Waals uniaxial ferrielectric $CuInP_2S_6$ as an example. We analyze the changes in the flexocoupling-induced phonon and ferron spectra arising from the appearance of spatially modulated phases induced by the flexoelectric coupling. We show that the free energy landscape of $CuInP_2S_6$ determines the specific features of its phonon spectra and ferron dispersion. We also discuss the contributions of optic and acoustic flexocoupling-induced ferrons to the pyroelectric and electrocaloric responses of $CuInP_2S_6$ at low temperatures.

[*] corresponding author, e-mail: anna.n.morozovska@gmail.com

[†] corresponding author, e-mail: vxg8@psu.edu

[‡] corresponding author, e-mail: lqc3@psu.edu

[§] corresponding author, e-mail: jhu238@wisc.edu

[**] corresponding author, e-mail: vysochanskii@gmail.com

## 1. INTRODUCTION

The static flexoelectric effect is a measure of the amount of electric polarization induced by a strain gradient and vice versa in solids [1]. There is ample evidence that the flexoelectric effect can make significant contributions to the electromechanics of meso- and especially nanoscale objects [2, 3], for which the strong strain gradients are inevitable present at the surfaces, interfaces, around point and topological defects [4, 5, 6]. In particular, the flexoelectric coupling (shortly “flexocoupling”) can play a decisive role in the emergence of unusual polar and conductive domain walls [7, 8] and spatially modulated phases [9] in ferroics and multiferroics, such as incipient and/or proper ferroelectrics, with or without antiferrodistortive long-range ordered phases. In fact, for several physical problems, such as the influence of the static [10] and dynamic [11] flexocouplings on the long-range order parameter fluctuations in the ordered phase of ferroics [12], the impact of flexocoupling can be critical. Existing experimental and theoretical results (see e.g., Refs. [13, 14, 15]) have been mostly focused on material-specific flexocoupling impacts on scattering spectra. Important, that experimental determination of phonon dispersion can be very informative to study the influence of flexocoupling on the spatially modulated long-range ordered phases in ferroelectrics. The basic experimental methods to obtain information about the fluctuations include dynamic dielectric measurements, neutron and Brillouin scattering [16, 17, 18].

The impact of flexocoupling on the generalized susceptibility as well as soft optical and acoustic phonon dispersions in the long-range ordered phases of ferroelectrics were previously studied within the framework of Landau-Ginzburg-Devonshire (LGD) approach in Ref. [19]. It was shown that the flexocoupling essentially broadens the momentum ($k$)-spectrum of generalized susceptibility and leads to the additional separation of the soft optical and acoustic mode phonon branches. Degeneration of the transverse optic and acoustic phonon modes disappears in the ferroelectric phase in comparison with the paraelectric phase due to the joint action of flexoelectric coupling and ferroelectric nonlinearity. It appeared that the dispersion of the soft optical phonons is less sensitive to the flexocoupling than that of acoustic phonons. In particular, the frequency of acoustic phonons tends to zero and becomes purely imaginary when the flexocoupling constant exceeds a critical value, which depends on the electrostriction and elastic constants, temperature, and gradient coefficients in the LGD free energy functional [20, 21]. When the frequency of the acoustic phonons goes to zero, a spatially modulated incommensurate phase may appear in commensurate ferroelectrics [21].

The concept of a “ferron” quasiparticle was introduced by Bauer et al. [22, 23] and Tang et al. [24], who identified the bosonic excitations in displacive ferroelectrics that carry electric dipoles from the phenomenological LGD approach and introduced the quasiparticles emerging from the joint action of anharmonicity and broken inversion symmetry in these materials. A ferron represents a collective amplitude mode of the long-range order parameter (spontaneous polarization) fluctuations in displacive ferroelectrics [22], including both volume-type [25] and surface-type ferron excitation modes, which carry electric dipoles [26, 27]. Compared to the ferroelectric soft optical phonons (soft mode), which are quanta of a coherent polarization wave with well-defined phase and frequency [28, 29], ferrons are typically not associated with a coherent wave and therefore they are incoherent. While the quanta of a coherent polarization wave are also referred to as coherent ferrons (as in Ref.[25]), the ferrons mentioned hereafter in this paper are all incoherent.

Using the calculated ferron spectrum in k-space, Tang et al. [24] predicted the significant contribution of ferrons to the temperature-dependent pyroelectric and electrocaloric properties at low temperatures, electric-field-tunable heat and polarization transport, and ferron-photon hybridization. Bauer et al. [25] argued that the introduction of dipole-carrying elementary incoherent excitations of the ferroelectric long-range order, i.e., ferrons, allows for the modeling of many observables and potentially leads to applications in thermal, information, and communication technologies. Subsequently, Wooten et al. [30] shows experimentally that an external electric field changes the velocity of the longitudinal acoustic ferrons, which carry heat and polarization, leading to the changes of thermal conductivity in a bulk ferroelectric. The revealed effect appeared four times larger than previously reported and was ascribed to the field-dependent scattering of phonons.

Recently the concept of ferron was extended by Yang and Chen [31] to the vector fluctuations of the long-range ordered spontaneous polarization, considering zero-point oscillations and thermal fluctuations. It was found that the finite temperature free energy of the long-range ordered polarization with bare parameters of the ground state (which can be found by the first-principles calculations) is renormalized by zero-point fluctuations at $T = 0$ K, and by thermal fluctuations at finite temperatures. The thermal excitations of the collective vector mode reduce the ferroelectric polarization to zero with heating and lead to the transition into nonpolar paraelectric phase at temperatures above $T_C$. Obviously, this occurs when fluctuations of polarization exceed the magnitude of the order parameter, namely the spontaneous polarization. Considering ferrons as vector fluctuations of the long-range ordered spontaneous polarization, the dimension of the vector space was taken equal to 3 for the case of displacive ferroelectrics like $PbTiO_3$ [31, 32]. For this case, Zhao et al. [32] have shown that the energy of ferrons is gapped in the long-wavelength limit at temperatures well below the ferroelectric-paraelectric phase transition temperature $T_C$; and the gap softens to minimal or gapless values as $T_C$ is approached, leading to a significant contribution to thermal properties.

A negative frequency of the unstable phonon mode, which exist in the incipient ferroelectrics (like $SrTiO_3$ and $KTaO_3$) according to the first principles calculations, is renormalized to a positive frequency by zero-point fluctuations in the ground state. The frequency of the long-wave ferrons, which are bosonic thermal excitations, increases proportionally to the thermally averaged polarization fluctuations $\langle \delta P^2(T) \rangle$. Based on this, the temperature dependence of the second power of the inverse dielectric permittivity, $\varepsilon^{-2}(T)$, is dominated by quantum fluctuations in the low temperature range. The temperature dependence of its first power, $\varepsilon^{-1}(T)$, is dominated by thermal fluctuations with heating [31]. The temperature dependence $\varepsilon^{-1}(T)$ of the quantum paraelectric $SrTiO_3$ shows an upturn below several kelvins instead of a plateau (at $T \leq T_s$) with cooling in the low temperature limit ($T \to 0$ K). This peculiarity can be explained by the electrostriction coupling of gapless acoustic phonons with polarization field [33]. The saturation temperature $T_s$ is related to the energy $E_0$ of zero-point fluctuations and crossover temperature $T_q$ as $k_B T_s = \frac{\hbar\omega_0}{4}$ and $T_s = \frac{T_q}{2}$ [34]. Note that the saturation temperature is well defined in the displacive limit involving one Einstein oscillator, while the crossover from a classical behavior to a saturated behavior becomes smeared (or even continuous) when several high-frequency phonon modes (which individually saturate at relatively high temperatures) are coupled to the phase transition order parameter [34]. In this case, the order parameter saturation appears at higher temperatures.

The main objective of this work is to extend concept of the "ferron" to the van der Waals (vdW) ferroelectrics [35]. For this purpose, we consider the layered (in fact 2D) vdW ferrielectric $CuInP_2S_6$ as an example because of its negative electrostriction [36] and the multi-well potential energy landscape [37, 38, 39, 40] leading to a range of interesting phenomena, including the unusual dynamics of polarization in $CuInP_2S_6$ thin films and nanoparticles [41, 42], the temperature and strain tunability of the multiple energy-degenerate metastable polar states in $CuInP_2S_6$ [43, 44], as well as to the emergence of controllable negative capacitance states [45, 46]. However, the contribution of ferrons to the phonon spectra and thermodynamic properties of $CuInP_2S_6$ has not been studied.

$CuInP_2S_6$ is a representee of a large family of phosphorous chalcogenides of metals, including 3D-ferroelectric $Sn_2P_2S_6$ and quantum paraelectric $Pb_2P_2S_6$ crystals [47]. The ferroelectric phase transition temperature $T_C$ tends to zero kelvin in the $(Pb_ySn_{1-y})_2P_2S_6$ compound at $y \to 0.61$. The crossover from quantum paraelectric behavior at low temperatures ($\varepsilon^{-1}(T) \sim T^2$) to classical Curie – Weiss behavior above the saturation temperature $T_s$ ($\varepsilon^{-1}(T) \sim T$) is observed in quantum paraelectric $Pb_2P_2S_6$ with $\omega_0 = 47$ cm$^{-1}$ and $E_0 = 0.003$ eV (72 K). For $Sn_2P_2S_6$, $\omega_0 = 60$ cm$^{-1}$ and $E_0 = 0.004$ eV (86 K). According to dielectric measurements [47], the crossover temperature $T_q$ is estimated as 190 K, proving the presence of several Einstein oscillators. It is worth noting that the energy $E_0$ correlates with Debye temperature $T_D$, which is about 85 K for $Pb_2P_2S_6$, and about 83 K for $Sn_2P_2S_6$ [48]. For the $CuInP_2S_6$ crystal, $T_D \approx 129$ K [49], and thus the ferrielectric phase transition at $T_C \approx 312$ K falls to the region of classic thermal fluctuations. The ferrielectric spontaneous polarization of $CuInP_2S_6$ contains the anticollinear contributions from $Cu^+$ and $In^{3+}$ cations moving out of the middle of the structural layers [50]. The $Cu^+$ cations obey ordering in the multi-well potential determined by the second order Jahn-Teller effect [51] in surrounding octahedron of sulfur atoms. The $In^{3+}$ cations demonstrate the displacement/ordering dynamics in a three-well local potential inside the sulfur octahedron [52]. Also, the sound velocity softening was observed by Brillouin scattering and ultrasonic measurements of $CuInP_2S_6$ crystals above $T_C$ [53, 54], which can be related to the presence of the long-range interactions among polar clusters and requires consideration of the flexoelectric coupling. Recent studies revealed that a very strong or "giant" flexoelectric effect determines domain engineering [55, 56], mechanical and optoelectronic properties [57, 58] of $CuInP_2S_6$.

To the best of our knowledge, theoretical studies of the dispersion of ferrons under flexocoupling (abbreviated as "*flexoferrons*" for brevity) and related polar, pyroelectric and thermal properties are currently missing. Here we calculate analytically the dispersion law $\omega(\boldsymbol{k})$ of soft optical and acoustic phonons, which refer to the ferroelectric soft mode and acoustic phonons under flexocoupling, respectively, and abbreviated as "*flexophonons*" for brevity, and flexoferrons of the vdW ferrielectric $CuInP_2S_6$. In addition to the flexocoupling, the effects of damping and higher elastic gradients are incorporated in these analytical calculations. The spectral density of flexoferrons, $\delta p(\boldsymbol{k})$, is also calculated. Our focus is on the changes in the flexophonon and flexoferron spectra related to the increase in the flexocoupling strength. We also analyze the contribution of optic and acoustic flexoferrons to the pyroelectric and electrocaloric responses of $CuInP_2S_6$ at low temperatures.

## 2. PROBLEM STATEMENT

Let us begin by deriving the dispersion relations of the soft optical phonon $\omega_O(\boldsymbol{k})$ (namely, the ferroelectric soft mode) and the acoustic phonon $\omega_A(\boldsymbol{k})$ (associated with the strain) by incorporating flexocoupling. In what follows, we name them the soft optical and acoustic flexophonons, respectively. According to the symmetry principle, the LGD free energy density $F$ should be invariant with respect to the symmetry transformations of the paraelectric phase. Using LGD theory in the considered case of $2/m$ point group symmetry (space group $C2/c$) of $CuInP_2S_6$ in the paraelectric phase, the Lagrange function $L = \int_t dt \int_{-\infty}^{\infty} d\boldsymbol{x}\,(F - K)$ consists of the kinetic energy $K$ and potential energy $F$. Following Ref. [19-21] the density of kinetic energy,

$$K = \frac{\mu}{2}\left(\frac{\partial P_3}{\partial t}\right)^2 + M\frac{\partial P_3}{\partial t}\frac{\partial U_3}{\partial t} + \frac{\rho}{2}\left(\frac{\partial U_3}{\partial t}\right)^2, \tag{1}$$

includes the dynamic flexocoupling with the coupling coefficient $M$; $\rho$ is the mass density of a material; the coefficient μ is the polarization inertia, which can be expressed via the vacuum dielectric constant $\varepsilon_0$ and the plasma frequency $\omega_p$ as $\mu = \frac{1}{\varepsilon_0\omega_p^2}$ [24]. Hereinafter we regard that $CuInP_2S_6$ as an uniaxial ferroelectric, whose polar axis is along the "3" direction, and so consider the coupling among the polarization component $P_3$, elastic displacement component $U_3$, and corresponding strains $u_{33}$, $u_{32}$ and $u_{31}$ (see **Scheme 1(a)**). The y-axis "2" is the second-order symmetry axis in the $CuInP_2S_6$ parent phase, which is normal to the monoclinic plane "$m$" in the ferrielectric phase.

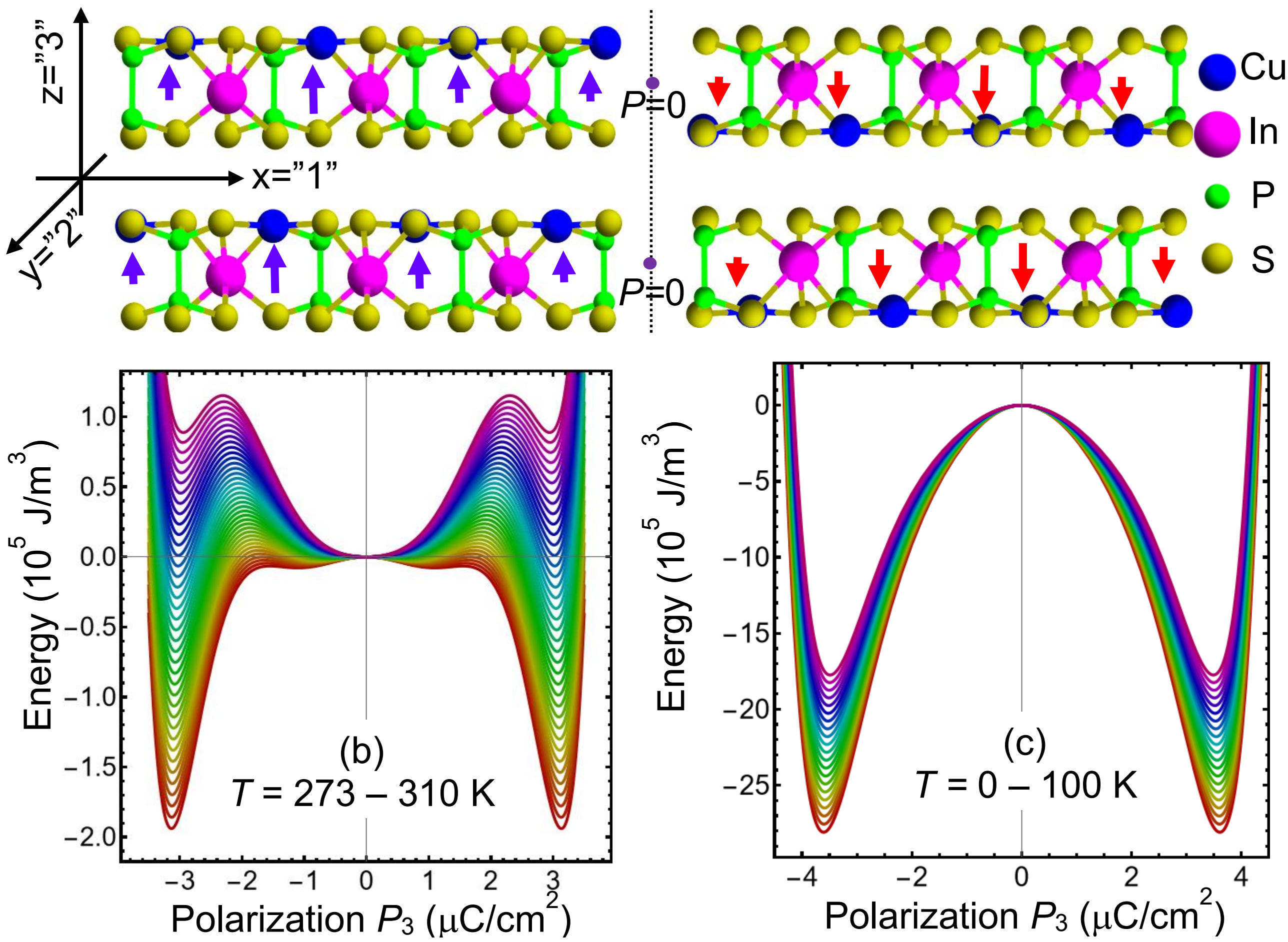

**SCHEME 1. (a)** Schematic of $CuInP_2S_6$ layers with crystallographic coordinate system. The polar axis $z$ ="3" is normal to the layers, and the axes $x$ ="1" and $y$ ="2" are in-plane of the layers. Blue and red arrows with varying length illustrate the transverse fluctuations of the ferrielectric polarization, $\delta P_3(x)$, with a wavevector **k** along "$x$". The vertical dotted line shows the break between the counter-polarized layers and corresponds to a virtual plane with zero polarization. **(b, c)** Schematic dependences of the $CuInP_2S_6$ free energy density on the polarization $P_3$ calculated for zero stresses and electric field. The curves color (from red to violet) corresponds to different temperatures $T$ from 273 K to 310 K with a step of 1 K **(b)** and from 0 K to 100 K with a step of 5 K **(c)**.

The bulk density $F$ of $CuInP_2S_6$ free energy as a function of $P_3$ and $u_{ij}$ and their gradients has the following form [43-46]:

$$F = F_{LGD} + F_{el+flexo}, \tag{2a}$$

$$F_{LGD} = \frac{\alpha}{2}P_3^2 + \frac{\beta}{4}P_3^4 + \frac{\gamma}{6}P_3^6 + \frac{\delta}{8}P_3^8 + g_{3i3j}\frac{\partial P_3}{\partial x_i}\frac{\partial P_3}{\partial x_j} - P_3E_3^{ext} - \frac{P_3E_3^d}{2}, \tag{2b}$$

$$F_{el+flexo} = -q_{ij33}u_{ij}P_3^2 - z_{ij33}u_{ij}P_3^4 + f_{3ijk}u_{jk}\frac{\partial P_3}{\partial x_i} + \frac{c_{ijkl}}{2}u_{ij}u_{kl} + \frac{v_{ijk}}{2}\left(\frac{\partial u_{ij}}{\partial x_k}\right)^2 - N_3 U_3. \quad (2c)$$

According to Landau theory [59, 60], the coefficient $\alpha$ depends linearly on the temperature $T$, $\alpha(T) = \alpha_T(T - T_C)$, which is valid for proper ferroelectrics well above quantum temperatures. The Barret-type expression, $\alpha(T) = \alpha_T T_q\left(\coth\frac{T_q}{T} - \coth\frac{T_q}{T_C}\right)$, where $T_C$ is the Curie temperature and $T_q$ is the quantum crossover temperature, is valid in a wider temperature range. All other coefficients in Eq.(2) are assumed to be weakly dependent on temperature. The Landau coefficient $\delta \geq 0$ for the stability of the free energy for all $P_3$ values. Note that the specific signs of Landau coefficients, namely $\alpha < 0$, $\beta > 0$, $\gamma < 0$, and $\delta > 0$ in the Landau expansion, $F_{Landau}(P_3) = \frac{\alpha}{2}P_3^2 + \frac{\beta}{4}P_3^4 + \frac{\gamma}{6}P_3^6 + \frac{\delta}{8}P_3^8$, make possible the appearance of the multi-well potential with lower and higher states of the spontaneous polarization in the definite temperature and/or strain range. For details see **Schemes 1(b)** and **1(c),** and **Table AI** in **Appendix A2** [61] for the values of Landau coefficients and their temperature dependences (collected from Refs.[41-46]).

Coefficients $g_{ijkl}$ are the polarization energy gradient coefficients. Coefficients $f_{ijkl}$ are the components of the static flexocoupling tensor. The coefficients $c_{ijkl}$ are the components of the elastic stiffness tensor. The coefficients, $q_{ijkl}$ and $z_{ijkl}$, are the second and higher order electrostriction coupling coefficients, respectively; $v_{ijk}$ are the coefficients of strain gradient energy. $N_3$ is the 3-component of the external mechanical force bulk density; $E_3^{ext}$ is the 3-rd component of external electric field. Note that the longitudinal fluctuations of polarization (i.e., $\delta P_3$) are much smaller than transverse fluctuations due to the depolarization field $E_3^d$ [62]. The field takes the form of $E_3^d = -\frac{\delta P_3}{\varepsilon_0 \varepsilon_b}$ in a thin $CuInP_2S_6$ membrane with infinitely large *xy* plane and contributes to the free energy by the term $-\frac{P_3 E_3^d}{2}$.

Considering the Tani mechanism and Khalatnikov relaxation, the explicit form of the LGD-KT equations is

$$\Lambda\frac{\partial U_3}{\partial t} = -\frac{\delta L}{\delta U_3}, \qquad \Gamma\frac{\partial P_3}{\partial t} = -\frac{\delta L}{\delta P_3}, \quad (3)$$

where $\Gamma$ and $\Lambda$ are phenomenological damping constants. Eqs.(3) linearized with respect to $P_3$ and $U_3$ allow calculating the generalized susceptibility of $CuInP_2S_6$ to the electric field and elastic force (see Eqs. (A.6) in **Appendix A2** [61]). The characteristic equation for the frequency dispersion, $\omega(\boldsymbol{k})$, of the flexophonons can be derived from the singularity of the generalized susceptibility in the Fourier $\{\boldsymbol{k}, \omega\}$-space.

Next, we derive and analyze the frequency dispersion of the flexoferrons at low temperatures, following the approach of Tang et al [24]. Generally, the electric dipole $p_i$ ($i$=1,2,3) carried by ferrons can be computed as $p_i = -\frac{\hbar\partial\omega_i}{\partial \boldsymbol{E}}$, where $\omega_i$ is the ferron frequency [25]. The frequency $\omega_i$ can be related to the frequences $\omega_A(\boldsymbol{k})$ and $\omega_O(\boldsymbol{k})$ calculated using Eqs. (1)-(3) above. In this regard, the flexoferrons can be classified into acoustic and optical flexoferrons, respectively. Specifically, we consider the case when the mechanical force $\boldsymbol{N}$ and the electric field $\boldsymbol{E}$ in Eqs. (3) are Langevin-type noise fields [24], which obey the fluctuation-dissipation theorem. Their correlators, averaged over the statistical ensemble in the "quantum" noise limit, have the following form in the Fourier $\{\boldsymbol{k}, \omega\}$-space [24]:

$$\langle \tilde{E}(\boldsymbol{k},\omega)\tilde{E}^*(\boldsymbol{k}',\omega')\rangle = \frac{\Gamma}{(2\pi)^2}\hbar\omega\coth\left(\frac{\hbar\omega}{2k_BT}\right)\delta(\boldsymbol{k}-\boldsymbol{k}')\delta(\omega-\omega'), \tag{4a}$$

$$\langle \tilde{N}(\boldsymbol{k},\omega)\tilde{N}^*(\boldsymbol{k}',\omega')\rangle = \frac{\Lambda}{(2\pi)^2}\hbar\omega\coth\left(\frac{\hbar\omega}{2k_BT}\right)\delta(\boldsymbol{k}-\boldsymbol{k}')\delta(\omega-\omega'). \tag{4b}$$

Here the tilda-symbol corresponds to the Fourier image in the $\{\boldsymbol{k}, \omega\}$-space. As anticipated $\langle \tilde{E}(\boldsymbol{k},\omega)\rangle = 0$ and $\langle \tilde{N}(\boldsymbol{k},\omega)\rangle = 0$. Assuming that the damping constant is small (namely $\left(\frac{\Gamma}{2\mu}\right)^2 \ll \omega_p^2$) and using the perturbation theory, in **Appendix A3** [61] we derived the second-order correction for the polarization response to the Langevin-type electric noise:

$$\langle p\rangle \approx \int_{-\infty}^{\infty}\frac{d^3\boldsymbol{k}}{(2\pi)^3}\left[\coth\left(\frac{\hbar\omega_A(\boldsymbol{k})}{2k_BT}\right)\delta p_A(\boldsymbol{k}) + \coth\left(\frac{\hbar\omega_O(\boldsymbol{k})}{2k_BT}\right)\delta p_O(\boldsymbol{k})\right], \tag{5}$$

where $\delta p_A(\boldsymbol{k})$ and $\delta p_O(\boldsymbol{k})$ are the "spectral densities" of acoustic and optical flexoferrons, $\omega_A(\boldsymbol{k})$ and $\omega_O(\boldsymbol{k})$ the eigen frequencies of optical and acoustic flexophonons calculated without damping (i.e., for $\Gamma = 0$, $\Lambda = 0$), which should be positive for the convergency of the integral in Eq.(5).

Pyroelectric response is the polarization variation in response to the temperature change, $P_3(T) = P_3(0) + \Delta P(T)$, where the contribution of the quantized fluctuations, represented by the flexoferrons, may become significant at temperatures much lower $T_q$ [24]. Following the approach proposed in Ref. [24], we postulate that the pyroelectric variation of polarization, $\Delta P(T)$, and the pyroelectric coefficient, $\Pi(T) = -\frac{d}{dT}\Delta P(T)$, are given by

$$\Delta P(T) = \int_{-\infty}^{\infty}\frac{d^3\boldsymbol{k}}{(2\pi)^3}\left(\frac{\delta p_O(\boldsymbol{k})}{\exp\left(\frac{\hbar\omega_O}{k_BT}\right)-1} + \frac{\delta p_A(\boldsymbol{k})}{\exp\left(\frac{\hbar\omega_A}{k_BT}\right)-1}\right), \tag{6a}$$

$$\Pi(T) = \int_{-\infty}^{\infty}\frac{d^3\boldsymbol{k}}{(2\pi)^3}\left(\frac{\exp\left(\frac{\hbar\omega_O}{k_BT}\right)\frac{\hbar\omega_O}{k_BT^2}}{\left[\exp\left(\frac{\hbar\omega_O}{k_BT}\right)-1\right]^2}\delta p_O(\boldsymbol{k}) + \frac{\exp\left(\frac{\hbar\omega_A}{k_BT}\right)\frac{\hbar\omega_A}{k_BT^2}}{\left[\exp\left(\frac{\hbar\omega_A}{k_BT}\right)-1\right]^2}\delta p_A(\boldsymbol{k})\right), \tag{6b}$$

where $\frac{1}{\exp\left(\frac{\hbar\omega}{k_BT}\right)-1}$ is the Boze-Einstein distribution function.

According to Maxwell relations, the electrocaloric (EC) coefficient, defined as the isothermal entropy change with electric field $E$, is equal to the pyroelectric effect, i.e., $\Sigma = \left(\frac{\partial\Delta S}{\partial E}\right)_T = \left(\frac{\partial\Delta P}{\partial T}\right)_E$. The experimentally observed temperature dependence of the EC response deviates strongly from a Curie-Weiss power-law. Using expressions for $\Pi(T, E)$, it is possible to estimate the EC response as [63]:

$$\Delta T_{EC}(E) \cong T\int_0^E \frac{1}{\rho_P C_P}\Pi(T,E)\,dE. \tag{7a}$$

$$\Sigma(E) \cong \Pi(T, E). \tag{7b}$$

Here $\rho_P$ is the mass density, $C_P$ is the heat capacity at constant pressure, and the field-dependent polarization $P(T, E)$ is substituted instead of the spontaneous polarization $P_S$. In **Appendix A4** [61] we calculated the contribution of the flexoferrons to pyroelectric and electrocaloric responses.

## 3. RESULTS AND DISCUSSION

### A. Analytical solutions for the eigen frequencies of the flexophonons and flexoferrons

In **Appendix A2** [61] we derived the characteristic equation for the frequency $\omega(\boldsymbol{k})$:

$$(\mu\rho - M^2)\omega^4 + i(\Gamma\rho + \Lambda\mu)\omega^3 - C(\boldsymbol{k})\omega^2 - i\omega\left[\Gamma\left(\hat{v}\boldsymbol{k}^4 + \hat{c}\boldsymbol{k}^2\right) + \Lambda(\alpha_S + \hat{g}\boldsymbol{k}^2)\right] + B(\boldsymbol{k}) = 0, \tag{8a}$$

where the functions $C(\boldsymbol{k})$ and $B(\boldsymbol{k})$ are:

$$C(\boldsymbol{k}) = \alpha_S\rho + \Gamma\Lambda + \left(\hat{c}\boldsymbol{k}^2\mu - 2\hat{f}\boldsymbol{k}^2 M + \hat{g}\boldsymbol{k}^2\rho\right) + \mu\hat{v}\boldsymbol{k}^4, \tag{8b}$$

$$B(\boldsymbol{k}) = \alpha_S\hat{c}\boldsymbol{k}^2 - 4P_S^2(\hat{q}\boldsymbol{k} + 2\hat{z}\boldsymbol{k}P_S^2)^2 + \hat{c}\boldsymbol{k}^2\hat{g}\boldsymbol{k}^2 - \left(\hat{f}\boldsymbol{k}^2\right)^2 + \alpha_S\hat{v}\boldsymbol{k}^4 + \hat{g}\boldsymbol{k}^2\hat{v}\boldsymbol{k}^4. \tag{8c}$$

Hereinafter a positive temperature-dependent function $\alpha_S$ is introduced as:

$$\alpha_S = \alpha + 3\beta P_S^2 + 5\gamma P_S^4 + 7\delta P_S^6 - 2q_{33}u_S, \tag{9}$$

where $P_S$ is the spontaneous polarization that corresponds to one of the stable potential wells (e.g., the deepest well) of the $CuInP_2S_6$ multi-well potential, and $u_S$ is the corresponding spontaneous strain. Since the temperature and strain behavior of $\alpha_S$ determines the specific features of the phonon and ferron dispersion; the dispersion features are also conditioned by the multi-well free energy landscape of $CuInP_2S_6$.

The tensorial convolutions $\hat{c}\boldsymbol{k}^2 = c_{3i3j}k_ik_j$, $\hat{v}\boldsymbol{k}^4 = v_{3ij3lm}k_ik_jk_lk_m$, $\hat{f}\boldsymbol{k}^2 = f_{3i3j}k_ik_j$, $\hat{g}\boldsymbol{k}^2 = g_{3i3j}k_ik_j + \frac{k_3^2}{\varepsilon_0\varepsilon_b k^2}$, $\hat{q}\boldsymbol{k} = g_{3i33}k_i$, and $\hat{z}\boldsymbol{k} = z_{3i33}k_i$ are detailed in **Appendix A2** [61] for the case of 2/m symmetry. Notably that the last term in $\hat{g}\boldsymbol{k}^2$, namely $\frac{k_3^2}{\varepsilon_0\varepsilon_b k^2}$, is related to the depolarization field $E_3^d$.

Equation (8) can be utilized to calculate the dispersion relation of optic and acoustic phonon modes, $\omega_O^{\mathrm{D}}(\boldsymbol{k})$ and $\omega_A^{\mathrm{D}}(\boldsymbol{k})$, respectively, where the superscript "D" is used to indicate that the damping factors $\Gamma$ and/or $\Lambda$. The dispersion relation is the same for the coherent and incoherent ferrons, and thus it is important to analyze the influence of the flexocoupling and damping on the dispersions of $\omega_O^{\mathrm{D}}(\boldsymbol{k})$ and $\omega_A^{\mathrm{D}}(\boldsymbol{k})$. In what follows, we set $k_2 = k_3 = 0$ for the discussions of the results, while deferring the detailed analysis of the depolarization effect (i.e., the case $k_3 \neq 0$) in different directions in the **k-**space to future studies. A general consideration should include the coupling between the ferrielectric polarization $P_3$ and all three components of displacement vector $U_i$, as required by the symmetry of the electrostrictive tensors $q_{ijkl}$ and $z_{ijkl}$ (see discussion after Eq. (2)). For the order-disorder type ferroelectrics Eqs.(8) allows to calculate the inverse relaxation time.

**Figures 1** and **2** show the dispersions of the real and imaginary parts of the optic and acoustic phonon modes frequencies, $\omega_O^{\mathrm{D}}(k_1)$ and $\omega_A^{\mathrm{D}}(k_1)$, for $\Lambda = 0$, $k_2 = k_3 = 0$, room ($T$ =293 K) and lower ($T$ =10 K) temperatures. The value of the damping constant $\Gamma$ varies for different curves, and the flexoelectric constant $f_{55}$ is fixed in **Fig. 1.** The flexoelectric coefficient $f_{55}$ varies for different curves and the damping coefficient $\Gamma$ is fixed in **Fig. 2**. Note that $\omega_A^{\mathrm{D}}$ and $\omega_O^{\mathrm{D}}$ are even functions of $k_1$.

The real part of $\omega_O^{\mathrm{D}}$ decreases relatively weakly with increase in $\Gamma$ (see **Fig.1(a)** and **1(c)**). The difference between the curves calculated for $\Gamma = 0$ and $\Gamma > 0$ is noticeable for $k_1 = 0$, slightly increases with increase in $k_1$, becomes maximal in the regions of optic and acoustic modes proximity (which is $0.25 < k_1 < 0.35$ nm$^{-1}$ for $T$ =293 K and $0.55 < k_1 < 0.65$ nm$^{-1}$ for $T$ =10 K) and then decreases with further increase in $k_1$. The real part of $\omega_A^{\mathrm{D}}$ is $\Gamma$-independent for small $k_1$. The dependence of $\mathrm{Re}[\omega_A^{\mathrm{D}}]$ on $\Gamma$ appears in the regions of the modes proximity, then its disappear in the cross-point of the $\mathrm{Im}[\omega_A^{\mathrm{D}}]$-curves (which is $k_1 \approx 0.3$ nm$^{-1}$ for $T$ =293 K and $k_1 \approx 0.6$ nm$^{-1}$ for $T$ =10 K) and appears again showing inverse tendency for further increase in $k_1$. The increase in $\Gamma$ leads to the so-called attraction of $\mathrm{Re}[\omega_O^{\mathrm{D}}]$ and $\mathrm{Re}[\omega_A^{\mathrm{D}}]$ dispersion curves, meaning that the distance between the curves becomes significantly smaller in the region of their proximity. The attraction is strongest at $k_1 \approx 0.3$ nm$^{-1}$ for $T$ =293 K and at $k_1 \approx 0.6$ nm$^{-1}$ for $T$ =10 K. The imaginary parts of $\omega_O^{\mathrm{D}}$ and $\omega_A^{\mathrm{D}}$ are zero for $\Gamma = 0$ and increase strongly with increase in $\Gamma$ (see **Fig.1(b)** and **1(d)**). The increase in $\Gamma$ leads to the significant difference between $\mathrm{Im}[\omega_O^{\mathrm{D}}]$ and $\mathrm{Im}[\omega_A^{\mathrm{D}}]$ curves, which is the strongest for the largest $\Gamma$ (namely for $\Gamma$ =0.015 s·m·J/C$^2$). However, the dispersion curves for $\mathrm{Im}[\omega_O^{\mathrm{D}}]$ and $\mathrm{Im}[\omega_A^{\mathrm{D}}]$ intersect near the point of the optic and acoustic modes proximity.

The real and imaginary parts of $\omega_O^{\mathrm{D}}$ and $\omega_A^{\mathrm{D}}$ are almost independent of the flexoelectric coupling for small $k_1$ (see **Fig.2(a)** and **2(c)**). When $k_1$ increases above 0.2 nm$^{-1}$ for $T$ =293 K (or above 0.4 nm$^{-1}$ for $T$ =10 K), $\mathrm{Re}[\omega_O^{\mathrm{D}}]$ begins to increase weakly and monotonically with increase in $f_{55}$. The difference between the dispersion curves of $\mathrm{Re}[\omega_O^{\mathrm{D}}]$ calculated for $f_{55} = 0$ and for $f_{55} > 0$ becomes maximal in the region of the optic and acoustic modes proximity. For $k_1 > 0.3$ nm$^{-1}$ and room temperature, $\mathrm{Re}[\omega_A^{\mathrm{D}}]$ decreases strongly with increase in $f_{55}$ and disappears for $k_1 = k_{cr}$, since $\omega_A^{\mathrm{D}}$ becomes purely imaginary $k_1 > k_{cr}$ (see red, orange and yellow curves of $\mathrm{Re}[\omega_A^{\mathrm{D}}]$ in **Fig.2(a)**). Note that the critical value $k_{cr}$ exists only for $f_{55} > f_{cr}(T)$, where $f_{cr} \approx 5$ V at 293 K, and it decreases with increase in $f_{55}$ (compare the ending of red, orange and yellow curves of $\mathrm{Re}[\omega_A^{\mathrm{D}}]$ in **Fig.2(a)**). At much lower temperatures (e.g., at $T$ =10 K) $\mathrm{Re}[\omega_A^{\Gamma}]$ decreases with increase in $k_1 > 0.5$ nm$^{-1}$ and becomes zero for very high $k_1 > 5$ nm$^{-1}$ and $f_{55} > f_{cr} > 12$ V, which are not shown in **Fig.2(c).** The dispersion curves for $\mathrm{Im}[\omega_O^{\mathrm{D}}]$ and $\mathrm{Im}[\omega_A^{\mathrm{D}}]$ intersect near the proximity point of the optic and acoustic modes for all values of $f_{55}$; and their dependence on $f_{55}$ becomes stronger with further increase in $k_1$ (see **Fig.2(b)** and **2(d)**).

It should be emphasized that the condition $\omega_A^{\mathrm{D}}(\boldsymbol{k},T)=0$ at nonzero $\boldsymbol{k}$ may correspond to the commensurate-incommensurate transition induced by the flexoelectric coupling [20, 21]. From Eqs.(8) the equation for the determination of $f_{cr}(\boldsymbol{k},T)$ is $B(\boldsymbol{k})=0$. For $CuInP_2S_6$ with 2/m symmetry and $k_2=k_3=0$, the explicit equation for $k_{cr}(f_{55},T)$ is:

$$g_{44}k_{cr}^4 v_{333}+\left(v_{333}\alpha_S-{f_{55}}^2+c_{55}g_{44}\right)k_{cr}^2+c_{55}\alpha_S-4P_S^2(q_{33}+2P_S^2 z_{33})^2=0. \quad (10a)$$

Here we used Voigt notation. From Eq.(10a), the analytical expressions for $k_{cr}$ and $f_{cr}$ are:

$$k_{cr}^2(f_{55},T)=\frac{{f_{55}}^2-v_{333}\alpha_S-c_{55}g_{44}\pm\sqrt{\left(v_{333}\alpha_S-{f_{55}}^2+c_{55}g_{44}\right)^2-4g_{44}v_{333}\left[c_{55}\alpha_S-4P_S^2\left(q_{33}+2P_S^2z_{33}\right)^2\right]}}{2g_{44}v_{333}}, \quad (10b)$$

$$f_{cr}^2(T)=v_{333}\alpha_S+c_{55}g_{44}-2\sqrt{g_{44}v_{333}[c_{55}\alpha_S-4P_S^2(q_{33}+2P_S^2z_{33})^2]}. \quad (10c)$$

Both signs "+" and "–" can have physical senses in Eq.(10b) when ${f_{55}}^2-v_{333}\alpha_S-c_{55}g_{44}>0$, meaning the "gap" of finite region in the acoustic phonon spectra. We chose the "–" sign before the radical in Eq.(10c) since $v_{333}$ is very small. Note that the condition $v_{333}=0$ leads to the expression $f_{cr}^2(T)=c_{55}g_{44}$, that agrees with the maximal values of the static flexoelectric effect coefficient (the "upper thermodynamic limit") established by Yudin, Ahluwalia and Tagantsev [10]. Expressions (10b)-(10c) describe analytically the anomalous behavior of $\mathrm{Re}[\omega_A^{\mathrm{D}}(k_1,f_{55})]$ induced by the high flexoelectric coupling constant (**Fig. 2(a)**).

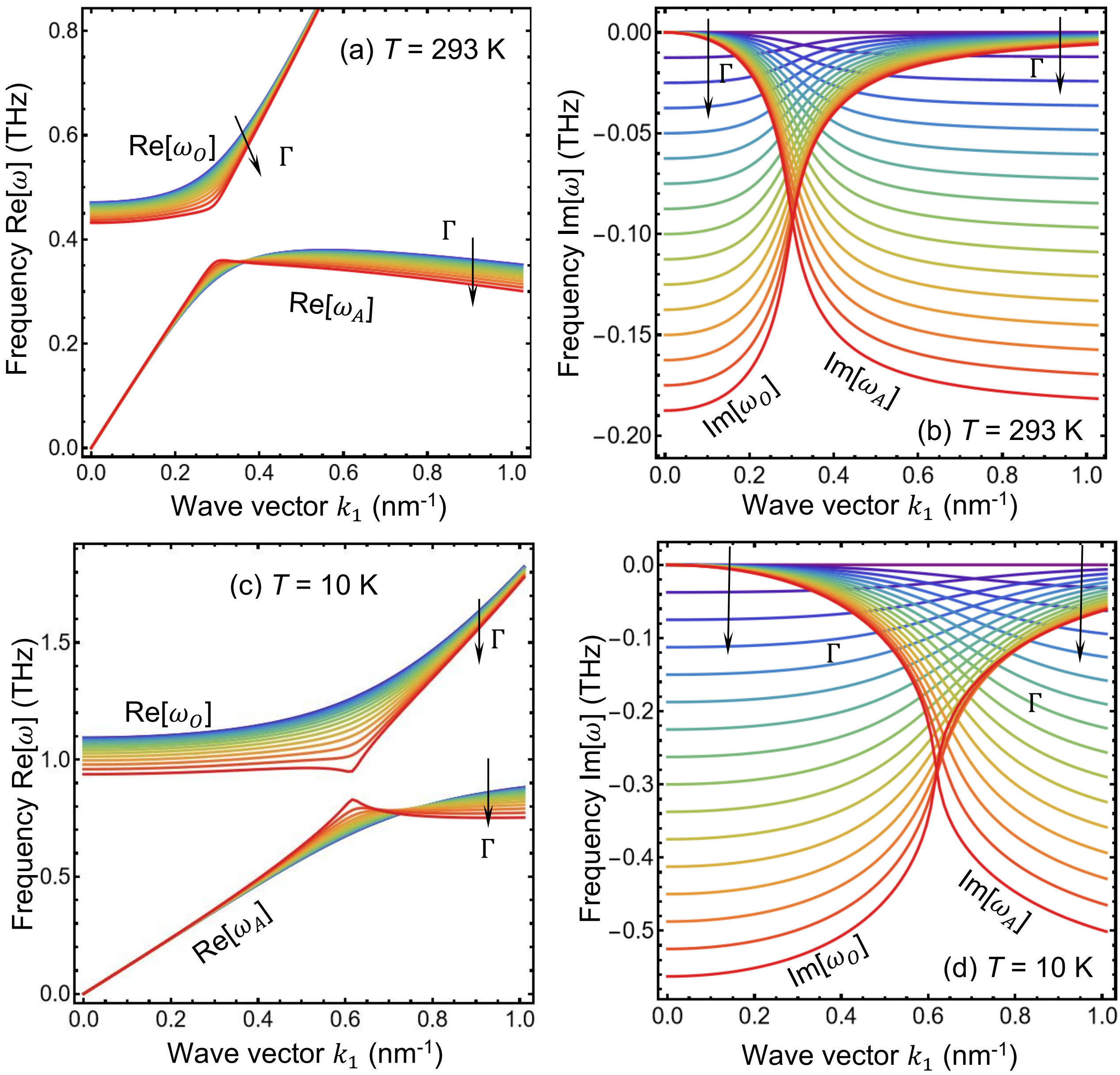

**FIGURE 1.** The dispersion of the real **(a, c)** and imaginary **(b, d)** parts of the optical and acoustic phonon modes, $\omega_O^D(k_1)$ and $\omega_A^D(k_1)$, calculated for $k_2 = k_3 = 0$, $T$ =293 K **(a, b)** and $T$ =10 K **(c, d)**. The value of damping constant Γ varies from 0 to 0.015 s·m·J/C$^2$ (with step size of 0.001 s·m·J/C$^2$) for different curves (from the dark blue to red color) in the plots **(a, b)**, and from 0 to 0.045 s·m·J/C$^2$ (with step size of 0.003 s·m·J/C$^2$) for different curves (from the dark blue to red color) in the plots **(c, d).** The flexoelectric coefficient $f_{55} = 5$ V and $\Lambda = 0$ for all curves. The material parameters for $CuInP_2S_6$ are listed in **Table AI**.

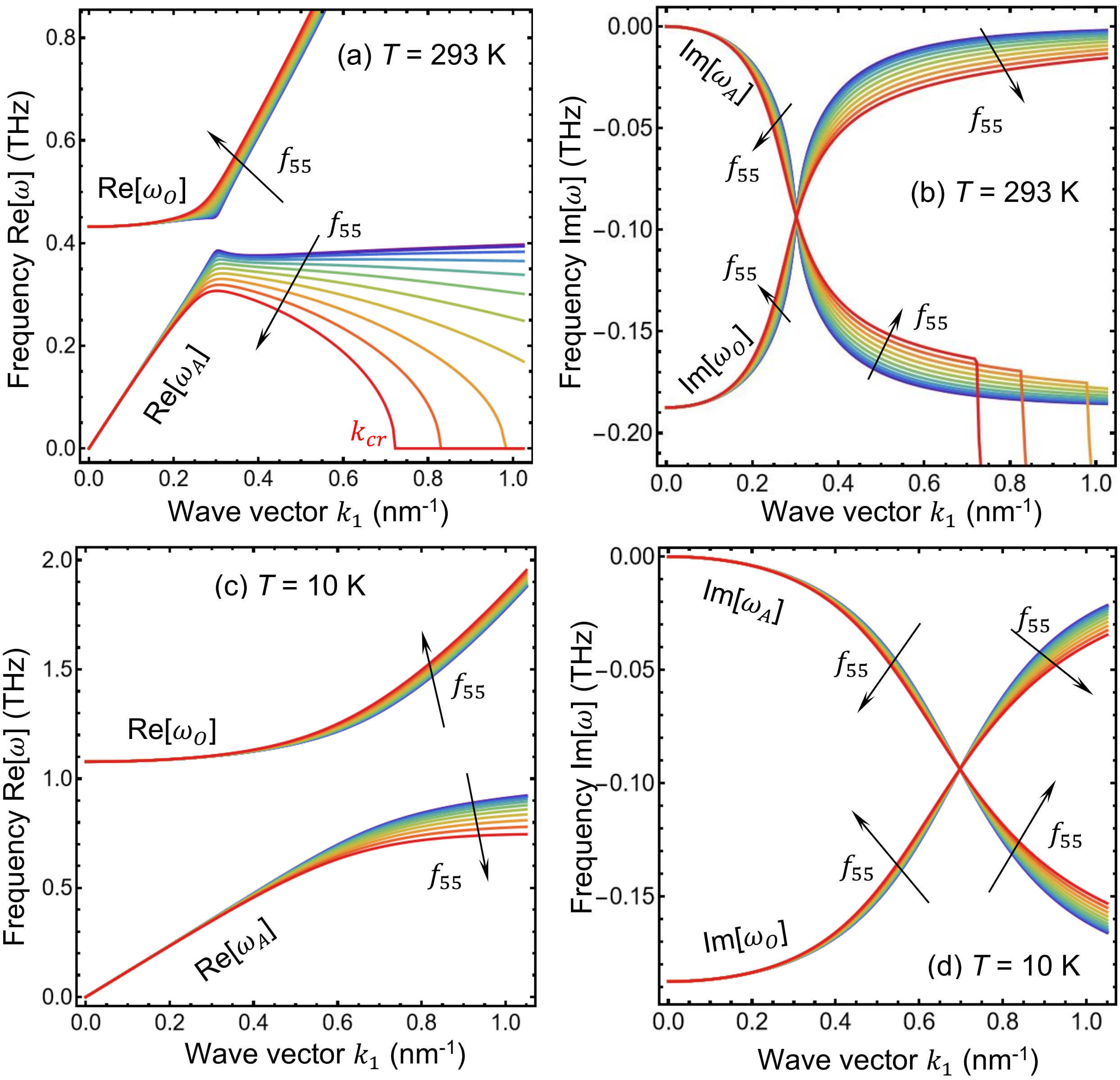

**FIGURE 2.** The dispersion of the real **(a, c)** and imaginary **(b, d)** parts of the optical and acoustic phonon modes, $\omega_O^D(k_1)$ and $\omega_A^D(k_1)$, calculated for $k_2 = k_3 = 0$, $T$ =293 K **(a, b)** and $T$ =10 K **(c, d)**. The flexoelectric coefficient $f_{55}$ varies from 1 to 10 V (with step size of 1 V) for different curves (from the dark blue to red color), the damping coefficients $\Gamma$ =0.015 s·m·J/C$^2$ and $\Lambda = 0$ for all curves. The material parameters of $CuInP_2S_6$ are listed in **Table AI**.

Without damping (i.e., at $\Gamma = 0$ and $\Lambda = 0$), the solution of Eq.(8a) can be represented in the form [19-21]:

$$\omega_{O,A}^2(\boldsymbol{k}) = \frac{C(\boldsymbol{k}) \pm \sqrt{C^2(\boldsymbol{k}) - 4(\mu\rho - M^2)B(\boldsymbol{k})}}{2(\mu\rho - M^2)}. \qquad (11)$$

Dispersion relation (11) contains optical (**O**) and acoustic (**A**) phonon modes, corresponding to the signs "+" and "–" before the radical, respectively. The "gap" between these modes is proportional to the value $\frac{\sqrt{C^2(\boldsymbol{k})-4(\mu\rho-M^2)B(\boldsymbol{k})}}{2(\mu\rho-M^2)}$.

The frequency $\omega_O$, given by Eq.(11), is a positive function that slightly increases with increase in $\boldsymbol{k}$ (see e.g., violet curves in **Fig. 1(a)** and **1(c)** corresponding to Re$[\omega_O]$ at $\Gamma = 0$). The dependence of $\omega_A$ on $\boldsymbol{k}$ becomes anomalous for high values of $f_{55}$, which is illustrated in **Fig. 3**.

**Figures 3(a)** and **3(b)** show the color maps of real and imaginary parts of the acoustic phonon mode frequency $\omega_A$ as a function of the wavenumber $k_1$ and flexoelectric coefficient $f_{55}$, calculated for $T = 293$ K and $T = 10$ K, respectively, and $k_2 = k_3 = 0$. The frequency $\omega_A$ is a monotonically increasing function of $\boldsymbol{k}$ for small $f_{55}$, then it decreases with further increase in $f_{55}$, becomes zero at $f_{55} = f_{cr}$ and purely imaginary for $f_{55} > f_{cr}$. The critical value of the flexoelectric coefficient $f_{cr}$ depends on temperature $T$ and wavevector $\boldsymbol{k}$. It is seen from **Figs. 3** that $f_{cr}$ increases strongly with increase in $T$. Also, $f_{cr}$ depends nonmonotonically on $k_1$: at first it decreases strongly with increase in $k_1$ from 0 to several nm$^{-1}$, reaches minimum, and then increases very slightly with further increase in $k_1$. Note that $\omega_A$ and $\omega_O$ are even functions of $k_1$.

The analytical expression for the function $f_{cr}(k_1, T)$ can be easily obtained from Eq.(10a):

$$f_{cr}(k_1, T) = \pm\sqrt{\frac{1}{k_1^2}[g_{44}k_1^4 v_{333} + c_{55}\alpha_S - 4P_S^2(q_{33} + 2P_S^2 z_{33})^2] - v_{333}\alpha_S - c_{55}g_{44}}. \qquad (12)$$

The expression (12) describes analytically the boundary $\omega_A(k_1, f_{55}) = 0$ shown in **Fig. 3.** As argued in Refs.[20-21], the instability occurring at $f_{55} > f_{cr}(\boldsymbol{k}, T)$ may indicate a transition into the spatially modulated incommensurate phase. Note that the features of the commensurate-incommensurate transition are observed in $CuInP_2S_6$ under high pressures [64]. However, the values of $f_{cr}$, corresponding to the possible transition, should be relatively high in accordance with our estimates (more than 6 – 12 V, see **Fig. 3**). At the same time the values of $f_{55}$ more than 4 – 6 V are higher than the thermodynamic limit of flexocoupling estimated in Ref.[10]. The estimate of some $f_{ij}$ obtained from current experiments [65] gives about 3 V, that is lower than the thermodynamic limit [10], but such moderate values unlikely can provide the giant flexoelectric effect observed by Chen et al. [55].

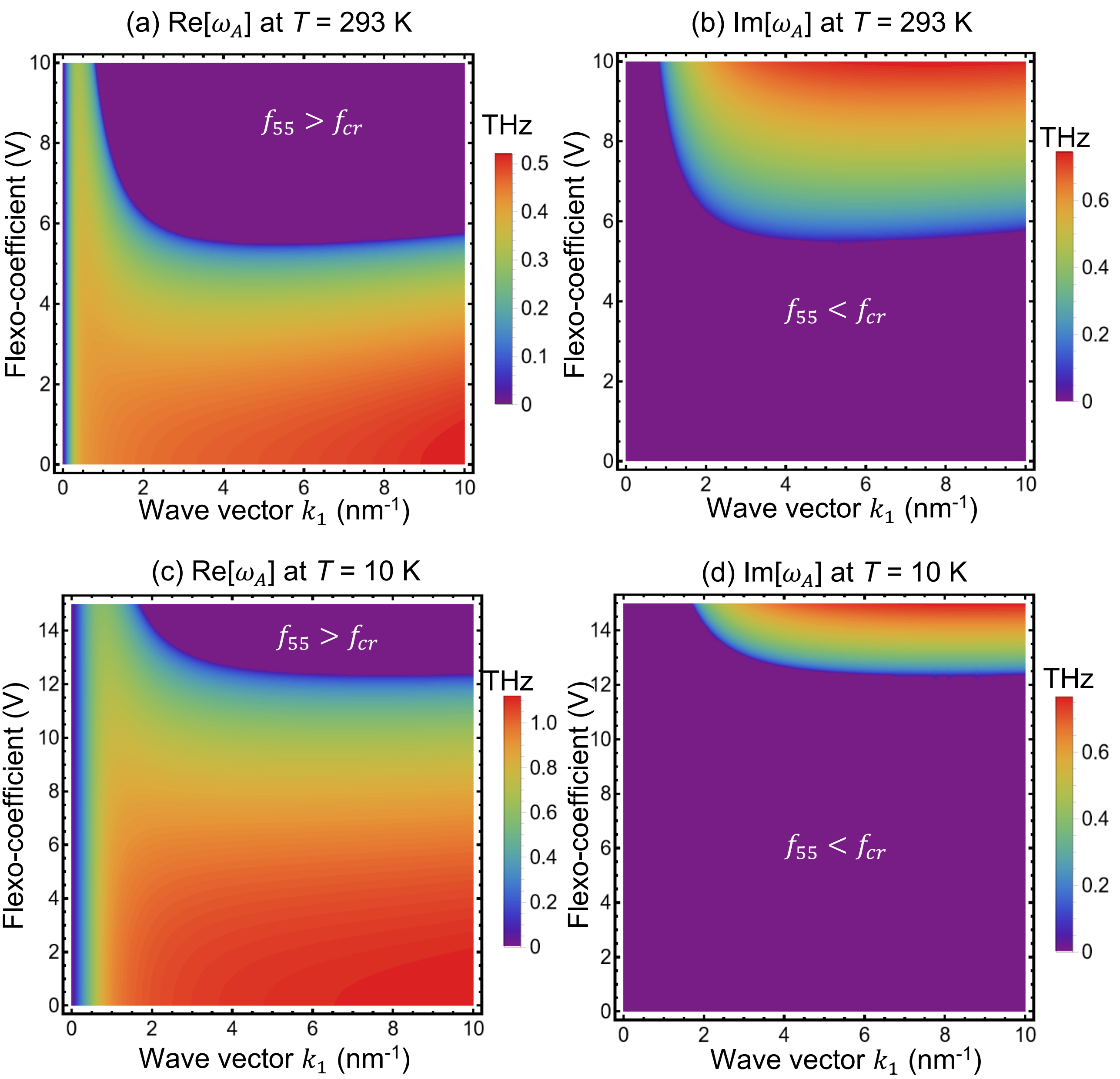

**FIGURE 3.** The color maps of the real **(a, c)** and imaginary **(b, d)** parts of the acoustic phonon mode frequency, $\omega_A$, as the function of the wavenumber $k_1$ and flexoelectric coefficient $f_{55}$ calculated for $k_2 = k_3 = 0$, $T$ =293 K **(a, b)** and 10 K **(c, d)**. The damping is absent ($\Gamma = 0, \Lambda = 0$) for all plots. The material parameters of $CuInP_2S_6$ are listed in **Table AI**.

### B. Analytical solution for the spectral density of flexoferrons

In **Appendix A3** [61] we derived approximate analytical expressions for the spectral densities of the acoustic and optic flexoferrons, $\delta p_A(\boldsymbol{k})$ and $\delta p_O(\boldsymbol{k})$:

$$\delta p_A(\boldsymbol{k}) \approx \frac{-\hbar}{2(\mu\rho - M^2)} \frac{3\beta^* P_S + 10\gamma^* P_S^3 + 21\delta P_S^5}{\alpha + 3\beta^* P_S^2 + 5\gamma^* P_S^4 + 7\delta P_S^6} \left| \frac{\hat{v}\boldsymbol{k}^4 + \hat{c}\boldsymbol{k}^2 - \rho\omega_A^2(\boldsymbol{k})}{\left(\omega_O^2(\boldsymbol{k}) - \omega_A^2(\boldsymbol{k})\right)\omega_A(\boldsymbol{k})} \right|, \tag{13a}$$

$$\delta p_O(\boldsymbol{k}) \approx \frac{-\hbar}{2(\mu\rho - M^2)} \frac{3\beta^* P_S + 10\gamma^* P_S^3 + 21\delta P_S^5}{\alpha + 3\beta^* P_S^2 + 5\gamma^* P_S^4 + 7\delta P_S^6} \left| \frac{\hat{v}\boldsymbol{k}^4 + \hat{c}\boldsymbol{k}^2 - \rho\omega_O^2(\boldsymbol{k})}{\left(\omega_A^2(\boldsymbol{k}) - \omega_O^2(\boldsymbol{k})\right)\omega_O(\boldsymbol{k})} \right|. \tag{13b}$$

Hereinafter the "effective" Landau coefficients, $\beta^*$ and $\gamma^*$ renormalized by the spontaneous strain are introduced, and expressions for $\omega_A(\boldsymbol{k})$ and $\omega_O(\boldsymbol{k})$ are given by Eqs.(11).

The k-dispersion of spectral densities $\delta p_A(k_1)$ and $\delta p_O(k_1)$ calculated from Eqs.(13) for $k_2 = k_3 = 0$, small damping (coefficient Γ less than 0.01 s·m·J/C$^2$), different values of the flexoelectric coefficient $f_{55}$, temperatures 10 K and 293 K are shown in **Fig. 4(a)** and **Fig. 4(b)**, respectively**.** Note that $\delta p_A$ and $\delta p_O$ are even functions of $k_1$.

The spectral density $\delta p_O$ is a regular and monotonic function of $k_1$ and $f_{55}$. Its magnitude depends weakly on $f_{55}$ and much more strongly depends on $k_1$ (see e.g., different curves for $\delta p_O$ in **Figs. 4(a)** and **4(b)**, as well as the color map in **Fig. A2**, **Appendix A3** [61]). The spectral density $\delta p_O$ is always negative. The absolute value $|\delta p_O|$ is maximal for $k_1 = 0$, and it monotonically decreases with increase in $k_1$ and tends to zero at $k_1 \to \infty$. Also $|\delta p_O|$ increases weakly with decrease in temperature. The intersection of $\delta p_O$ and $\delta p_A$ curves happen very close to the $k_1$-point of the frequencies $\omega_O(k_1)$ and $\omega_A(k_1)$ maximal proximity.

The spectral density $\delta p_A$ is a regular negative function of $k_1$ at $f_{55} < f_{cr}$. The function monotonically decreases with increase in $k_1$ and saturates for high $k_1$ (see e.g., all $\delta p_A$-curves in **Figs. 4(a)**, blue, green and yellow curves $\delta p_A$-curves in **Figs. 4(b)**). For $f_{55} > f_{cr}$ $\delta p_A$-curves becomes a strongly nonmonotonic function of $k_1$ that diverge ($\delta p_A \to -\infty$) at $k_1 = k_{cr}(f_{55}, T)$. The position of the $\delta p_A$ divergency is determined by the condition $\omega_A(k_1, T) = 0$, because $\delta p_A(k_1) \sim \frac{1}{\omega_A(k_1)}$ in accordance with Eq.(13a).

The color maps of $\delta p_A$ as the function of the wavenumber $k_1$ and flexoelectric coefficient $f_{55}$ are shown in **Fig. 4(c)** and **4(d)**, for $k_2 = k_3 = 0$, at $T$ =10 K and $T$ =293 K, respectively. Since dynamic instability of acoustic mode occurs for $f_{55} \geq f_{cr}(k_1, T)$, the spectral density of acoustic flexoferron diverges at $f_{55} = f_{cr}(k_1, T)$ (compare **Fig. 4(c)** and **4(d)** with the behavior of $\omega_A(k_1, T)$ shown in **Fig. 3**). It is seen from **Figs. 4(c)** and **4(d)** that the width of the range where $|\delta p_A|$ reaches high values (e.g., $|\delta p_A| > 1$) decreases slightly with increase in $T$; and the position of the divergency curve ($\delta p_A \to -\infty$) shifts down with increase in $T$. The divergency curve nonmonotonically depends on $k_1$: at first it decreases strongly with increase in $k_1$ from 0 to several nm$^{-1}$, reaches minimum, and then increases very slightly with further increase in $k_1$.

Since $\delta p_A(\boldsymbol{k})$ diverges at $\omega_A(\boldsymbol{k}) = 0$ and $f_{55} = f_{cr}$, as well as $\omega_A(k)$ becomes imaginary when $f_{55} > f_{cr}$, the acoustic contribution to the integral in the approximate expression (5) also diverges, representing the fingerprint of a dynamic instability induced by the flexoelectric coupling (e.g., possible transition to a spatially modulated incommensurate phase). Thus one should consider the case $f_{55} < f_{cr}$ for correct integration in Eq.(5). The case $f_{55} > f_{cr}$, when $f_{55}$ is above the thermodynamic limit [10], requires other approaches for the correct analytical calculations of the flexophonon and flexoferron contributions to the polarization response.

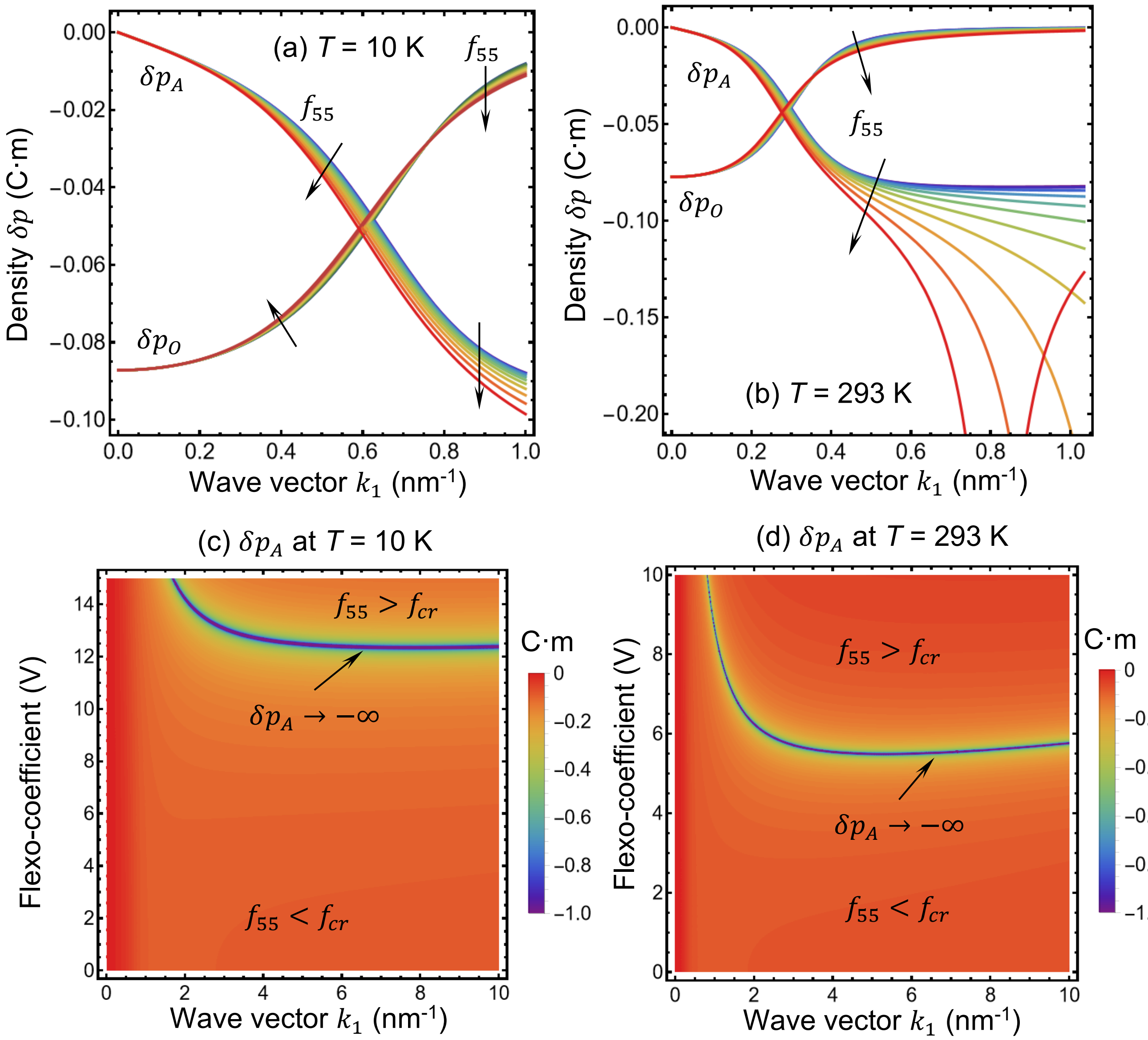

**FIGURE 4.** The k-dispersion of flexoferron spectral densities $\delta p_A(k_1)$ and $\delta p_O(k_1)$ (in $10^{-30}$ C·m) calculated for $k_2 = k_3 = 0$, temperature $T = 10$ K **(a)** and 293 K **(b)**, and different values of the flexoelectric coefficient $f_{55}$ varying from 0 to 10 V with step size of 1 V for different curves. The color maps of the $\delta p_A(k_1)$ as a function of the wavenumber $k_1$ and flexoelectric coefficient $f_{55}$ calculated for $k_2 = k_3 = 0$, $T = 10$ K **(c)** and $T = 293$ K **(d)**. The material parameters of $CuInP_2S_6$ are listed in **Table AI**.

### C. The contribution of flexoferrons to pyroelectric and electrocaloric responses

Temperature dependences of the polarization variation $\Delta P$ and pyroelectric coefficient $\Pi$ calculated from Eq. (6) for low temperatures, fixed damping coefficient $\Gamma$, and different values of the flexoelectric coefficient $f_{55}$ are shown in **Fig. 5(a)** and **5(b)**, respectively. The temperature variation of polarization and pyroelectric coefficients increase weakly with increase in the flexoelectric coefficient; also they increase with increase in temperature according to the power law. The saturation of the pyroelectric response with temperature may appear at the temperatures comparable or higher than the crossover temperature $T_q \cong 50$ K. The saturation is almost independent of the value of $f_{55}$. Temperature dependences of the acoustic and optic flexoferron contributions to the pyroelectric coefficient Π calculated for different values of the flexoelectric coefficient $f_{55}$ are shown in **Fig. A3** in **Appendix A4** [61]. It is seen from the figure that the contribution of the optic ferrons becomes much lower (in 2 – 6 orders of magnitude) than the contribution of the acoustic ferrons for the temperatures above (0.5 – 1) K.

Electric field dependences of the polarization variation $\Delta P$ and pyroelectric coefficient Π calculated from Eq.(6) for different values of the flexoelectric coefficient $f_{55}$ and $T$ =10 K are shown in **Fig. 5(c)** and **5(d)**, respectively**.** These dependences reflect the contribution of the flexoferrons to the electrocaloric response. In this case the contribution of the acoustic ferrons dominates over the contribution of the optic ferrons by 5 – 6 orders of magnitude. The termination of the curves in **Fig. 5(c)** and **5(d)** correspond to the critical value of the field for which the acoustic flexoferron frequency becomes zero. The critical field is much lower than the thermodynamic coercive field of $CuInP_2S_6$ at 10 K (~ 1.3 MV/cm). Notably that the critical value of the electric field exists at $f_{55} > f_{cr}$ and depends significantly on the flexoelectric coefficient $f_{55}$: it decreases strongly with increase in $f_{55}$. We would like to underline that the dependence of the ferron spectra on the applied electric field agrees with earlier results of Wooten et al. [30].

To explain numerical results, one should derive approximate, analytical expressions for the pyroelectric response. In **Appendix A4** [61], we derive the approximate expression for the pyroelectric charge $\Delta Q$ using the method of the steepest descent for the integration with exponential functions in Eqs.(6),

$$\Delta Q(T) \approx \frac{-\hbar\rho}{2(\mu\rho-M^2)}\frac{3\beta^* P_S+10\gamma^* P_S^3+21\delta P_S^5}{\alpha+3\beta^* P_S^2+5\gamma^* P_S^4+7\delta P_S^6}\left[\frac{1}{\omega_O}\sqrt{\frac{2\pi k_B T}{\hbar\omega_O''}}\exp\left(-\frac{\hbar\omega_O}{k_B T}\right)+\frac{12\,v}{\omega_O^2}\left(\frac{\rho}{c}\right)^2\left(\frac{k_B T}{\hbar}\right)^4\right]. \quad (14)$$

Hereinafter $\omega_O \equiv \omega_O(0)$ and $\omega_O'' = \frac{d^2}{dk^2}\omega_O(k)\Big|_{k\to 0}$.

Disregarding the temperature dependence of the LGD coefficients at low temperatures, the temperature derivative of the pyroelectric charge, $q(T) = \frac{Q(T)}{dT}$, can be estimated from Eq.(14) as:

$$q(T) \approx \frac{\hbar\rho}{2(\mu\rho-M^2)}\frac{3\beta^* P_S+10\gamma^* P_S^3+21\delta P_S^5}{\alpha+3\beta^* P_S^2+5\gamma^* P_S^4+7\delta P_S^6}\left[\left(\frac{1}{\omega_O}\sqrt{\frac{\pi k_B}{2\hbar\omega_O'' T}}+\sqrt{\frac{2\hbar\pi}{k_B\omega_O'' T^3}}\right)\exp\left(-\frac{\hbar\omega_O}{k_B T}\right)+\frac{12\,v}{\omega_O^2}\left(\frac{k_B}{\hbar}\right)^4\left(\frac{\rho}{c}\right)^2 T^3\right]. \quad (15)$$

Expressions (14)-(15) are valid only at low temperatures and do not consider the role of the flexoelectric coupling correctly. However, they explain the dominant contribution of the acoustic ferrons to the pyroelectric and electrocaloric responses in comparison to the contribution of the optic ferrons which exponentially vanishes at low temperatures.

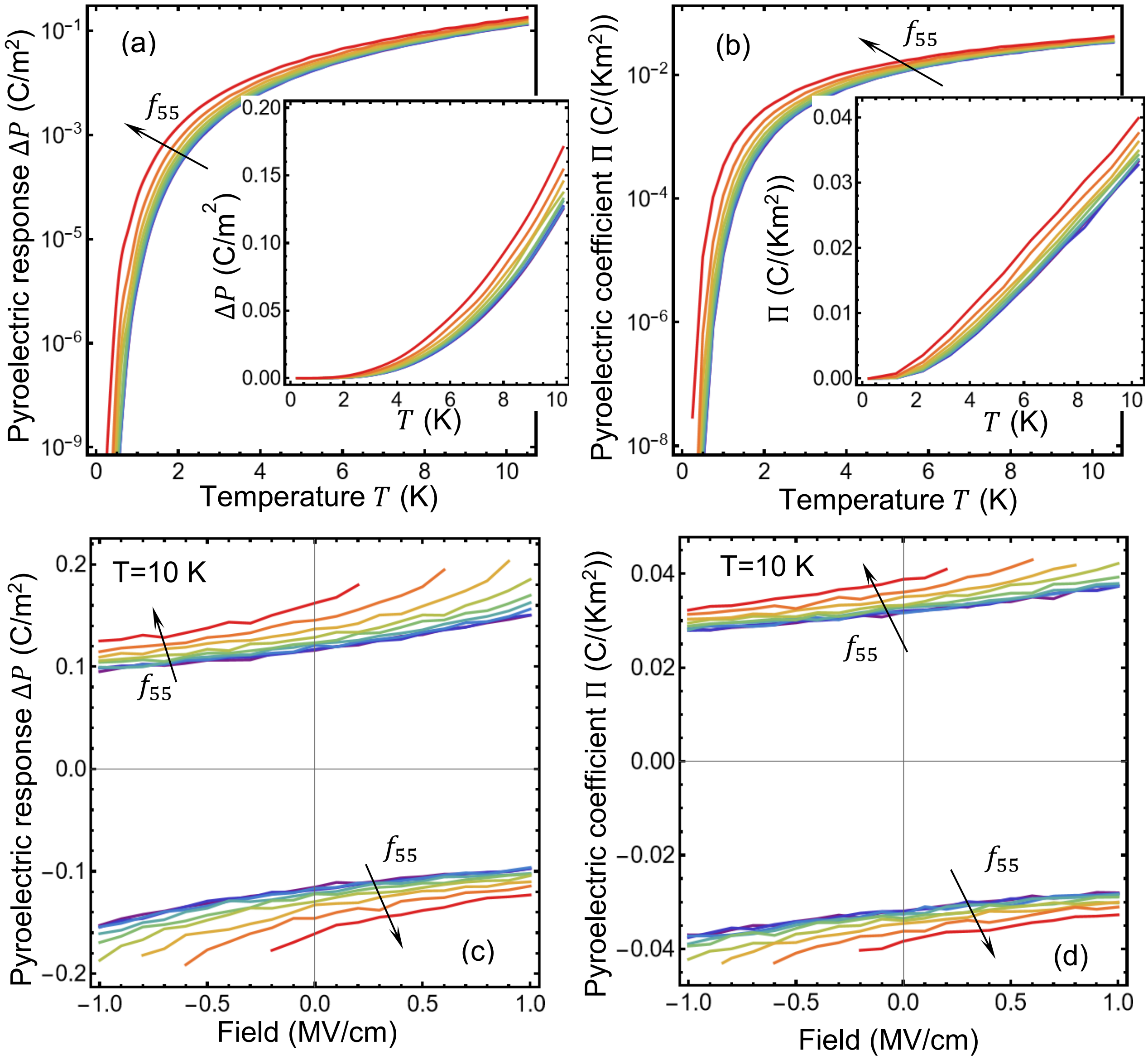

**FIGURE 5.** Temperature **(a, b)** and electric field **(c, d)** dependences of the polarization variation $\Delta P$ **(a, c)** and pyroelectric coefficient $\Pi$ **(b, d)** calculated from Eq.(6) for different values of the flexoelectric coefficient $f_{55}$, varying from 0 to 8 V with the step of 1 V (see different curves). Insets in the plots (a) and (b) show the dependences $\Delta P(T)$ and $\Pi(T)$ in a linear scale. $CuInP_2S_6$ material parameters are listed in **Table AI**, the damping is neglected ($\Gamma = \Lambda = 0$).

Polarization variation $\Delta P$ and pyroelectric coefficient $\Pi$ as the function of flexoelectric coefficient $f_{55}$ and electric field $E_3$ calculated for $T = 10$ K and without damping ($\Gamma = \Lambda = 0$) are shown in **Fig. 6(a)** and **6(b)**, respectively. It is seen that the increase of the flexocoupling constant $f_{55}$ leads to a significant increase in $\Delta P$ and $\Pi$; at that these values become much less sensitive to the magnitude of the applied electric field due to the increase of the internal flexoelectric field, which is proportional to the $f_{55}$. It is seen that the contour lines of $\Delta P$ and $\Pi$ have almost the same shape, which is corroborated by the analytical expressions (14)-(15). From these expressions $\Pi \sim \Delta P/T$ at low temperatures.

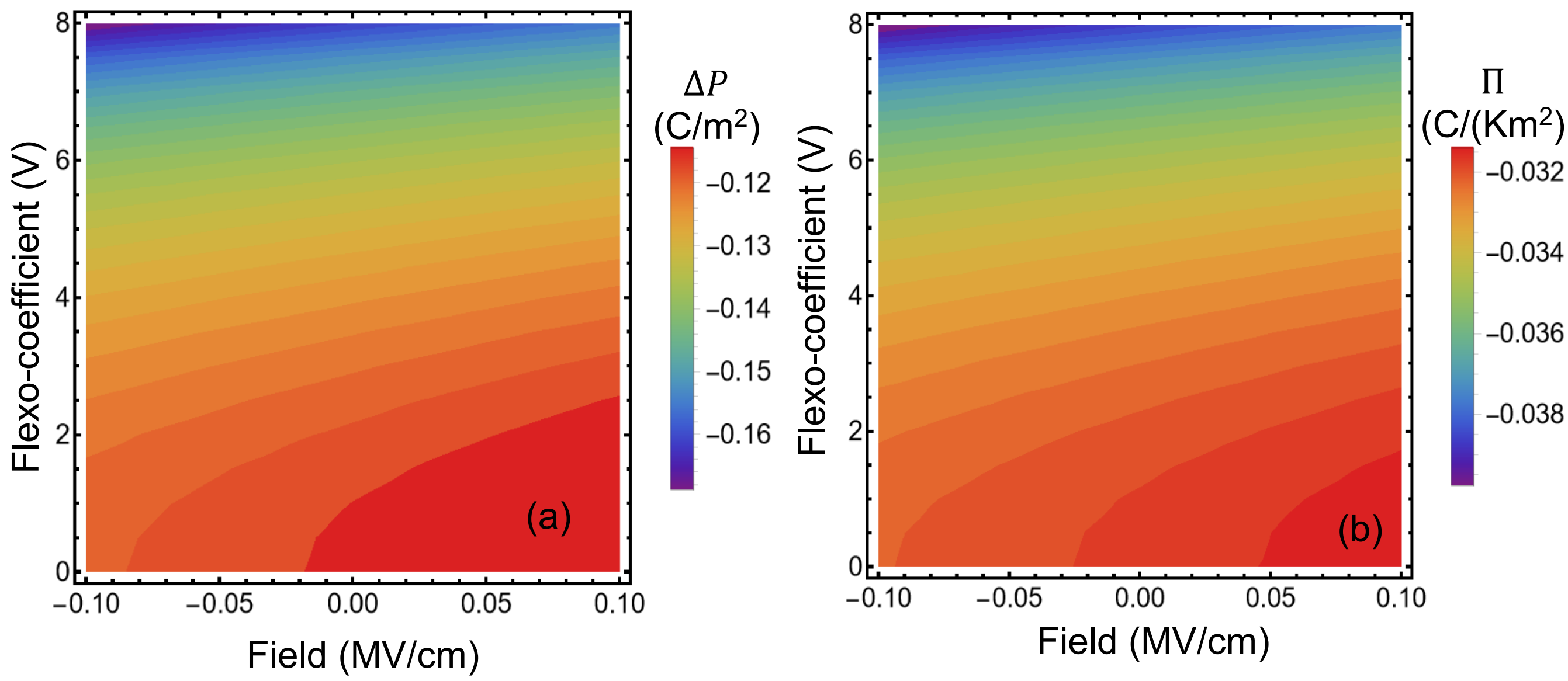

**FIGURE 6.** Color maps of polarization vatiation $\Delta P$ **(a)** and pyroelectric coefficient $\Pi$ **(b)** as a function of flexoelectric coefficient $f_{55}$ and electric field $E_3$. The maps are calculated using Eq.(6) for $CuInP_2S_6$ material parameters (listed in **Table AI**), $T = 10$ K and without damping ($\Gamma = \Lambda = 0$).

## 4. CONCLUSION

We analytically obtain the dispersion law of soft optical and acoustic flexophonons and flexoferrons by incorporating flexocoupling, damping, and higher elastic gradients in the Landau-Ginzburg-Devonshire free energy functional of the vdW uniaxial ferrielectric $CuInP_2S_6$.

The increase of the flexocoupling constant $f_{55}$ leads to dramatic changes in the flexophonon and flexoferron spectra related to possible appearance of spatially modulated phases induced by the flexoelectric coupling [20, 21]. In particular, the acoustic flexophonon frequency $\omega_A^{\mathrm{D}}(\boldsymbol{k}, T)$ becomes zero at $f_{55} = f_{cr}(k_1, T)$ and purely imaginary at $f_{55} > f_{cr}(k_1, T)$. The condition $\omega_A^{\mathrm{D}}(\boldsymbol{k}, T) = 0$ at nonzero $\boldsymbol{k}$ corresponds to the dynamic instability induced by the increase in the flexoelectric coupling magnitude, which may indicate the commensurate-incommensurate transition. However, the observability of the transition is questionable, because the magnitude of $f_{cr}$ can be higher than the thermodynamic limit estimated in Ref.[10]. Since the dynamic instability occurs for $f_{55} > f_{cr}(k_1, T)$, the behavior of acoustic flexoferron responses to the transition at $f_{55} = f_{cr}(k_1, T)$, where the spectral density $\delta p_A(\boldsymbol{k})$ diverges.

The contribution of acoustic flexoferrons to the pyroelectric and electrocaloric responses of $CuInP_2S_6$ at low temperatures dominates over that of the optic flexoferrons by several orders of magnitude. The performed calculations of the electrocaloric response make physical sense for electric fields below the critical value. The critical field corresponds to the situation when the frequency of acoustic flexoferron becomes zero (which is possible only for the case $f_{55} > f_{cr}$). The critical value of the electric field decreases strongly with increase in the flexoelectric coefficient $f_{55}$. Notably that the dependence of the ferron spectra on applied electric field agrees with earlier results of Wooten et al. [30].

To verify theoretical predictions made in this work, inelastic neutron scattering measurements of phonon dispersion performed in a wide temperature range (e.g., at low and room temperatures) in $CuInP_2S_6$ seem urgent. Analysis of the neutron scattering results together with available Raman spectroscopy [37, 64] and ultrasound measurements [53, 54] results will allow to estimate the static and dynamic flexocoupling constants $f_{55}$ and M, as well as to determine the damping constants Γ and Λ. Next, low temperature measurements of pyroelectric and electrocaloric responses of $CuInP_2S_6$ layers with various degrees of bending can be suggested to observe the flexoferron contribution.

**Authors' contribution.** A.N.M. generated the research idea, formulated the problem, performed most of the analytical calculations and wrote the manuscript draft jointly with Y.M.V. E.A.E. with O.V.B. and M.Y.Y. wrote the codes and prepared figures. Y.Z., J-M.H., G.-D.Z., V.G., L.-Q. C. and Y.M.V. worked on the analysis of results and manuscript improvement.

**Acknowledgements.** Authors are very grateful to the Referees for fruitful discussions and valuable suggestions. The work of A.N.M., E.A.E., G.-D.Z., Y.Z., V.G., J-M.H., and L.-Q.C. is supported by the DOE Software Project on "Computational Mesoscale Science and Open Software for Quantum Materials", under Award Number DE-SC0020145 as part of the Computational Materials Sciences Program of US Department of Energy, Office of Science, Basic Energy Sciences. The work of O.V.B. is sponsored by the Target Program of the National Academy of Sciences of Ukraine, Project No. 5.8/25-П "Energy-saving and environmentally friendly nanoscale ferroics for the development of sensorics, nanoelectronics and spintronics". Y.M.V. and A.N.M. also acknowledge support from the Horizon Europe Framework Programme (HORIZON-TMA-MSCA-SE), project № 101131229, Piezoelectricity in 2D-materials: materials, modeling, and applications (PIEZO 2D). Results were visualized in Mathematica 14.0 [66].

# Supplementary Materials to
# Flexocoupling-induced phonons and ferrons in van der Waals ferroelectrics

Anna N. Morozovska[1††], Eugene. A. Eliseev[2], Oleksiy V. Bereznikov[1], Mykola Ye. Yelisieiev[3], Guo-Dong Zhao[4], Yujie Zhu[5], Venkatraman Gopalan[4‡‡], Long-Qing Chen[4§§], Jia-Mian Hu[5***], and Yulian M. Vysochanskii[6†††]

[1] Institute of Physics, National Academy of Sciences of Ukraine, 46, pr. Nauky, 03028 Kyiv, Ukraine

[2] Frantsevich Institute for Problems in Materials Science, National Academy of Sciences of Ukraine, Omeliana Pritsaka str., 3, Kyiv, 03142, Ukraine

[3] Institute of Semiconductor Physics, National Academy of Sciences of Ukraine, 45, pr. Nauky, 03028 Kyiv, Ukraine

[4] Department of Materials Science and Engineering, Pennsylvania State University, University Park, PA 16802, USA

[5] Department of Materials Science and Engineering, University of Wisconsin-Madison, Madison, WI, 53706, USA

[6] Institute of Solid-State Physics and Chemistry, Uzhhorod University, 88000 Uzhhorod, Ukraine

## APPENDIX A. Calculation details

### A1. The free energy functional

Paraelectric and ferroelectric phases of $CuInP_2S_6$ have the point symmetry $2/m$ and $m$, respectively. It was shown that in V-d-W ferroelectric $CuInP_2S_6$, the axis 2 lies in the plane of layers, with spontaneous polarization pointed along the normal vector to the layers. The coordinate system is organized as follows, the coordinate axis "$X_2$" is along the symmetry axis "2", the coordinate axis "$X_3$" is along the normal vector of all the layers, while the coordinate axis "$X_1$" is perpendicular to "$X_3$" and $X_2$". Note that the mirror plane (the only symmetry element of the ferroelectric phase) is parallel to the axes "$X_1$" and "$X_3$". According to the symmetry principle the Landau free energy should be invariant with respect to the symmetry transformations of the paraelectric phase, so that the non-zero tensorial elements may have the even number of indices "2" and/or odd number of "1" and "3", but the total number of both "1" and "3" should be even also.

[††] corresponding author, e-mail: anna.n.morozovska@gmail.com

[‡‡] corresponding author, e-mail: vxg8@psu.edu

[§§] corresponding author, e-mail: lqc3@psu.edu

[***] corresponding author, e-mail: jhu238@wisc.edu

[†††] corresponding author, e-mail: vysochanskii@gmail.com

Using LGD theory and scalar approximation in the considered one-component 1D case, the Lagrange function $L = \int_t dt \int_{-\infty}^{\infty} dx\,(F - K)$ consists of the kinetic energy K and free energy $F$ of ferroelectric. The density of kinetic energy,

$$K = \frac{\mu}{2}\left(\frac{\partial P_3}{\partial t}\right)^2 + M\frac{\partial P_3}{\partial t}\frac{\partial U_3}{\partial t} + \frac{\rho}{2}\left(\frac{\partial U_3}{\partial t}\right)^2, \qquad \text{(A.1)}$$

includes the dynamic flexocoupling with the magnitude $M$; ρ is the mass density of a material; the coefficient μ is the polarization inertia, which can be expressed via the vacuum dielectric constant $\varepsilon_0$ and the plasma frequency $\omega_p$ as $\mu = \frac{1}{\varepsilon_0 \omega_p^2}$. Hereinafter we regard that $CuInP_2S_6$ is a uniaxial ferroelectric, which polar axis is "3", and so consider the coupling between the polarization component $P_3$, elastic displacement component $U_3$, and corresponding strains $u_3$, $u_4$ and $u_5$ (in Voigt notations).

The bulk density of the free energy $F$ that depends on polarization component $P_3$ and strain component $u$, and their gradients, has the following form:

$$F = \frac{\alpha}{2}P_3^2 + \frac{\beta}{4}P_3^4 + \frac{\gamma}{6}P_3^6 + \frac{\delta}{8}P_3^8 + \frac{g_{55}}{2}\left(\frac{\partial P_3}{\partial x_1}\right)^2 + \frac{g_{44}}{2}\left(\frac{\partial P_3}{\partial x_2}\right)^2 + g_{35}\frac{\partial P_3}{\partial x_3}\frac{\partial P_3}{\partial x_1} + \frac{g_{33}}{2}\left(\frac{\partial P_3}{\partial x_3}\right)^2 - P_3 E_3^{ext} - \frac{P_3 E_3^d}{2} +$$
$$- q_{53}u_5 P_3^2 - q_{33}u_3 P_3^2 - z_{533}u_5 P_3^4 - z_{333}u_3 P_3^4 +$$
$$+ f_{55}u_5\frac{\partial P_3}{\partial x_1} + f_{53}u_5\frac{\partial P_3}{\partial x_3} + f_{44}u_4\frac{\partial P_3}{\partial x_2} + f_{33}u_3\frac{\partial P_3}{\partial x_3} + f_{35}u_3\frac{\partial P_3}{\partial x_1} +$$
$$+ \frac{c_{55}}{2}u_5^2 + \frac{c_{44}}{2}u_4^2 + c_{35}u_5 u_3 + \frac{c_{33}}{2}u_3^2 + \frac{v_{5151}}{2}\left(\frac{\partial u_5}{\partial x_1}\right)^2 + \frac{v_{4242}}{2}\left(\frac{\partial u_4}{\partial x_2}\right)^2 + \frac{v_{3333}}{2}\left(\frac{\partial u_3}{\partial x_3}\right)^2 - N_3 U_3 \qquad \text{(A.2)}$$

Note that $q_{43} \equiv 0$ and $z_{433} \equiv 0$ for the considered symmetry group. According to Landau theory, the coefficient α linearly depends on the temperature $T$ for proper ferroelectrics, $\alpha(T) = \alpha_T(T - T_C)$, which is valid well above quantum temperatures. The Barret-type expression, $\alpha(T) = \alpha_T T_q\left(\coth\frac{T_q}{T} - \coth\frac{T_q}{T_C}\right)$, where $T_C$ is the Curie temperature and $T_q$ is the quantum vibration temperature, is valid in a wide temperature range (from law to high temperatures). All other coefficients in Eq.(A.2) are supposed to be temperature independent. The coefficient $\delta \geq 0$ for the stability of the free energy for all $P$ values. The gradient coefficients $g_{ij}$ determines the magnitude of the gradient energy. Coefficients $f_{ij}$ are the components of the static flexocoupling tensor. Coefficients $c_{ii}$ are elastic stiffness. The coefficients $q_{ij}$ and $z_{ijk}$ are second-order and higher-order electrostriction coupling coefficients, respectively. $N_3$ is z-component of the external mechanical force bulk density; $E_3^{ext}$ is z-component of external electric field. Note that the longitudinal fluctuations of polarization are much smaller due to the depolarization field $E_3^d$, which contribution to the free energy is given by the term $\frac{P_3 E_3^d}{2}$.

## A2. Analytical solution for the eigen frequency of the optical and acoustic flexophonons

Considering the Tani mechanism and Khalatnikov relaxation, the explicit form of the LGD-KT equations, $\delta L/\delta U_3 = 0$ and $\delta L/\delta P_3 = 0$, is:

$$v_{3ij3kl}\frac{\partial^4 U_3}{\partial x_i \partial x_j \partial x_k \partial x_l} + \rho\frac{\partial^2 U_3}{\partial t^2} + \Lambda\frac{\partial U_3}{\partial t} - c_{55}\frac{\partial^2 U_3}{\partial x_1^2} - c_{44}\frac{\partial^2 U_3}{\partial x_2^2} - 2c_{35}\frac{\partial^2 U_3}{\partial x_1 \partial x_3} - c_{33}\frac{\partial^2 U_3}{\partial x_3^2} - f_{55}\frac{\partial^2 P_3}{\partial x_1^2} - f_{44}\frac{\partial^2 P_3}{\partial x_2^2} - (f_{35} + f_{53})\frac{\partial^2 P_3}{\partial x_1 \partial x_3} - f_{33}\frac{\partial^2 P_3}{\partial x_3^2} + 2P_3\left(q_{53}\frac{\partial P_3}{\partial x_1} + q_{33}\frac{\partial P_3}{\partial x_3}\right) + 4\left(z_{533}\frac{\partial P_3}{\partial x_1} + z_{333}\frac{\partial P_3}{\partial x_3}\right)P_3^3 + M\frac{\partial^2 P_3}{\partial t^2} = N. \qquad \text{(A.3a)}$$

$$\mu\frac{\partial^2 P_3}{\partial t^2} + \Gamma\frac{\partial P_3}{\partial t} + \alpha P_3 + \beta P_3^3 + \gamma P_3^5 + \delta P_3^7 - g_{55}\frac{\partial^2 P_3}{\partial x_1^2} - g_{44}\frac{\partial^2 P_3}{\partial x_2^2} - 2g_{35}\frac{\partial^2 P_3}{\partial x_1 \partial x_3} - g_{33}\frac{\partial^2 P_3}{\partial x_3^2} - f_{55}\frac{\partial^2 U_3}{\partial x_1^2} - f_{44}\frac{\partial^2 U_3}{\partial x_2^2} - (f_{35} + f_{53})\frac{\partial^2 U_3}{\partial x_1 \partial x_3} - f_{33}\frac{\partial^2 U_3}{\partial x_3^2} - 2\left(q_{53}\frac{\partial U_3}{\partial x_1} + q_{33}\frac{\partial U_3}{\partial x_3}\right)P_3 - 4\left(z_{533}\frac{\partial U_3}{\partial x_1} + z_{333}\frac{\partial U_3}{\partial x_3}\right)P_3^3 + M\frac{\partial^2 U_3}{\partial t^2} = E_3^{ext} + E_3^d, \qquad \text{(A.3b)}$$

where $\Gamma$ and $\Lambda$ are phenomenological damping constants.

Next, let use the Fourier integral expansions for polarization $P$, displacement $U$, perturbation electric field $E$ and mechanical force density $N$:

$$P = P_S + \int d\,\omega \int d^3\boldsymbol{k}\, exp(i\boldsymbol{kx} - i\omega t)\,\tilde{P}, \quad U = u_{S3j}x_j + \int d\,\omega \int d^3\boldsymbol{k}\, exp(i\boldsymbol{kx} - i\omega t)\,\tilde{U}, \qquad \text{(A.4a)}$$

$$E = \int d\,\omega \int d^3\boldsymbol{k}\, exp(i\boldsymbol{kx} - i\omega t)\,\tilde{E}, \qquad N = \int d\,\omega \int d^3\boldsymbol{k}\, exp(i\boldsymbol{kx} - i\omega t)\,\tilde{N}. \qquad \text{(A.4b)}$$

Hereinafter $P_S$ is the spontaneous polarization $u_S = \frac{1}{c}(qP_S^2 + zP_S^4)$ is the spontaneous strain. For the case $\delta = 0, \gamma > 0$ the analytical expression $P_S^2 = \left(\sqrt{\beta^2 - 4\alpha\gamma} - \beta\right)/2\gamma$ is valid.

In the Fourier representation, linearized Eqs.(A.3) have the form:

$$\left(\hat{v}\boldsymbol{k}^4 + \hat{c}\boldsymbol{k}^2 - i\Lambda\omega - \rho\omega^2\right)\tilde{U} + \left(\hat{f}\boldsymbol{k}^2 + 2i\hat{q}\boldsymbol{k}P_S + 4i\hat{z}\boldsymbol{k}P_S^3 - M\omega^2\right)\tilde{P} = \tilde{N}, \qquad \text{(A.5a)}$$

$$(-i\Gamma\omega - \mu\omega^2 + \alpha_S + \hat{g}\boldsymbol{k}^2)\tilde{P} + \left(\hat{f}\boldsymbol{k}^2 - 2i\hat{q}\boldsymbol{k}P_S - 4i\hat{z}\boldsymbol{k}P_S^3 - M\omega^2\right)\tilde{U} = \tilde{E}. \qquad \text{(A.5b)}$$

Hereinafter the designations are introduced:

$$\hat{c}\boldsymbol{k}^2 \stackrel{\text{def}}{=} c_{55}k_1^2 + 2c_{53}k_1k_3 + c_{44}k_2^2 + c_{33}k_3^2, \qquad \text{(A.5c)}$$

$$\hat{v}\boldsymbol{k}^4 \stackrel{\text{def}}{=} v_{3ij3lm}k_ik_jk_lk_m = v_{5151}k_1^4 + 2v_{4233}k_1^2k_3^2 + v_{4242}k_2^4 + v_{3333}k_3^4, \qquad \text{(A.5d)}$$

$$\hat{f}\boldsymbol{k}^2 \stackrel{\text{def}}{=} f_{55}k_1^2 + f_{44}k_2^2 + (f_{35} + f_{35})k_1k_3 + f_{33}k_3^2 \qquad \text{(A.5e)}$$

$$\hat{g}\boldsymbol{k}^2 \stackrel{\text{def}}{=} g_{55}k_1^2 + g_{44}k_2^2 + 2\mathrm{g}_{35}k_1k_3 + g_{33}k_3^2 + \frac{k_3^2}{\varepsilon_0\varepsilon_b k^2}, \qquad \text{(A.5f)}$$

$$\hat{q}\boldsymbol{k} \stackrel{\text{def}}{=} q_{53}k_1 + q_{33}k_3, \; \hat{z}\boldsymbol{k} \stackrel{\text{def}}{=} z_{533}k_1 + z_{333}k_3, \qquad \text{(A.5g)}$$

$$k^2 \stackrel{\text{def}}{=} k_1^2 + k_2^2 + k_3^2. \qquad \text{(A.5h)}$$

The positive temperature-dependent function $\alpha_S$:

$$\alpha_S = \alpha + 3\beta P_S^2 + 5\gamma P_S^4 + 7\delta P_S^6 - 2q_{33}u_S \equiv \alpha + 3\beta^* P_S^2 + 5\gamma^* P_S^4 + 7\delta P_S^6. \qquad \text{(A.5i)}$$

Hereinafter the "effective" Landau coefficients, $\beta^*$ and $\gamma^*$, which are renormalized by the spontaneous strain, are introduced. In the scalar case $\beta^* = \beta - \frac{2}{3}\frac{q^2}{c}$ and $\gamma^* = \gamma - \frac{2}{5}\frac{z^2}{c}$. The solution of Eqs.(A.5a)-(A.5b) has the matrix form

$$\begin{pmatrix} \tilde{P} \\ \tilde{U} \end{pmatrix} = \begin{pmatrix} \tilde{\chi}(\boldsymbol{k},\omega) & \tilde{\eta}(\boldsymbol{k},\omega) \\ \tilde{\eta}^*(\boldsymbol{k},\omega) & \tilde{\vartheta}(\boldsymbol{k},\omega) \end{pmatrix} \begin{pmatrix} \tilde{E} \\ \tilde{N} \end{pmatrix}, \tag{A.6a}$$

where the matrix elements, which are generalized susceptibilities, are given by expressions:

$$\tilde{\chi}(\boldsymbol{k},\omega) = \frac{\hat{v}\boldsymbol{k}^4 + \hat{c}\boldsymbol{k}^2 - \rho\omega^2 - i\Lambda\omega}{\Delta(\boldsymbol{k},\omega)}, \quad \tilde{\eta}(\boldsymbol{k},\omega) = -\frac{\hat{f}\boldsymbol{k}^2 - M\omega^2 - 2i\hat{q}\boldsymbol{k}P_S - 4i\hat{q}\boldsymbol{k}P_S^3}{\Delta(\boldsymbol{k},\omega)}, \quad \tilde{\vartheta}(\boldsymbol{k},\omega) = \frac{\alpha_S + \hat{g}\boldsymbol{k}^2 - \mu\omega^2 - i\Gamma\omega}{\Delta(\boldsymbol{k},\omega)}, \tag{A.6b}$$

$$\Delta(\boldsymbol{k},\omega) = (\alpha_S + \hat{g}\boldsymbol{k}^2 - \mu\omega^2 - i\Gamma\omega)\left(\hat{v}\boldsymbol{k}^4 + \hat{c}\boldsymbol{k}^2 - \rho\omega^2 - i\Lambda\omega\right) - \left(\hat{f}\boldsymbol{k}^2 - M\omega^2\right)^2 - 4k^2P_S^2(\hat{q}\boldsymbol{k} + 2\hat{z}\boldsymbol{k}P_S^2)^2. \tag{A.6c}$$

From the singularity of the generalized susceptibility corresponding to $\Delta(k,\omega) = 0$ one derives the characteristic equation for the frequency $\omega(\boldsymbol{k})$, which form is the power expansion on $\omega$, is the following:

$$(\mu\rho - M^2)\omega^4 + i(\Gamma\rho + \Lambda\mu)\omega^3 - C(\boldsymbol{k})\omega^2 - i\omega\left[\Gamma\left(\hat{v}\boldsymbol{k}^4 + \hat{c}\boldsymbol{k}^2\right) + \Lambda(\alpha_S + \hat{g}\boldsymbol{k}^2)\right] + B(\boldsymbol{k}) = 0, \tag{A.7a}$$

where the functions $C(\boldsymbol{k})$ and $B(\boldsymbol{k})$ are introduced:

$$C(\boldsymbol{k}) = \alpha_S\rho + \Gamma\Lambda + \left(\hat{c}\boldsymbol{k}^2\mu - 2\hat{f}\boldsymbol{k}^2M + \hat{g}\boldsymbol{k}^2\rho\right) + \mu\hat{v}\boldsymbol{k}^4, \tag{A.7b}$$

$$B(\boldsymbol{k}) = \alpha_S\hat{c}\boldsymbol{k}^2 - 4P_S^2(\hat{q}\boldsymbol{k} + 2\hat{z}\boldsymbol{k}P_S^2)^2 + \hat{c}\boldsymbol{k}^2\hat{g}\boldsymbol{k}^2 + \alpha_S\hat{v}\boldsymbol{k}^4 - \left(\hat{f}\boldsymbol{k}^2\right)^2 + \hat{g}\boldsymbol{k}^2\hat{v}\boldsymbol{k}^4. \tag{A.7c}$$

Without damping (i.e. at $\Gamma = \Lambda = 0$), the solution of biquadratic Eq.(A.7a) can be represented in the form:

$$\omega_{O,A}^2(\boldsymbol{k}) = \frac{C(\boldsymbol{k}) \pm \sqrt{C^2(\boldsymbol{k}) - 4(\mu\rho - M^2)B(\boldsymbol{k})}}{2(\mu\rho - M^2)}, \tag{A.8}$$

Dispersion relation (A.8) contains one optical (**O**) and one (**A**) phonon modes, which corresponds to the signs "+" and "–" before the radical, respectively.

The dispersion of the real and imaginary parts of the optic and acoustic phonon modes frequencies, $\omega_O^{\mathrm{D}}(k_1)$ and $\omega_A^{\mathrm{D}}(k_1)$, calculated for $\Lambda = 0$ and $k_2 = k_3 = 0$ are shown in **Figs. 1** and **2** in the main text. The material parameters of $CuInP_2S_6$ are listed in **Table AI**. Notably that the last term in Eq.(A.5f), $\frac{k_3^2}{\varepsilon_0\varepsilon_b k^2}$, is related with the depolarization field. The depolarization field leads to almost total suppression of longitudinal phonons. Hence the assumption $k_3 = 0$ used in **Figs. 1-3** seems grounded.

**Table AI.** LGD parameters for a bulk ferrielectric $CuInP_2S_6$ Helmholtz free energy $F(P_3, u_{ij})$ with $P_3$ and $u_{ij}$ as independent variables (collected from Refs.[18-20, 41-45] listed in the main text)

| Parameter (dimensionality) | Value |
|---|---|
| $\varepsilon_b$ | 9 |
| $\alpha_T$ ($C^{-2}$·m J/K) | $1.64067\times10^{7}$ |
| $T_{C,q}$ (K) | $T_C \cong 292.67$, $T_q \cong 50$ |
| $\beta$ ($C^{-4}$·$m^5$J) | $8.43\times10^{12}(1.-0.00239\,T+2.28\times10^{-6}T^2)$ |
| $\gamma$ ($C^{-6}$·$m^9$J) | $-1.67283\times10^{16}(1-0.00249\,T+3.389\times10^{-6}\,T^2)$ |
| $\delta$ ($C^{-8}$·$m^{13}$J) | $9.824\times10^{18}(1.-0.00127\,T+4.0999\times10^{-6}T^2)$ |
| $q_{i3}$ (J $C^{-2}$·m) | $q_{13}=1.4879\times10^{11}(1-0.00206\,T)$<br>$q_{23}=1.0603\times10^{11}(1-0.00203\,T)$<br>$q_{33}=-4.0334\times10^{11}(1-0.00188\,T)$<br>$q_{53}=-7.3*10^{10}$ |
| $z_{i33}$ ($C^{-4}$·$m^5$J) * | $z_{133}=-1.414\times10^{14}(1-0.00099\,T)$<br>$z_{233}=-0.774\times10^{14}(1-0.00146\,T)$<br>$z_{333}=1.181\times10^{14}(1-0.00699\,T)$<br>$z_{533}=10^{13}$ |
| $s_{ij}$ ($Pa^{-1}$) **<br>$c_{ij}$ (Pa) | $s_{11}=1.510\times10^{-11}$ , $s_{12}=0.183\times10^{-11}$**<br>$c_{11}=6.803\times10^{10}$ , $c_{12}=-7.364\times10^{9}$** |
| $g_{3i3j}$ (J $m^3/C^2$)*** | Estimated parameter, which has an order of $10^{-10}$, e.g., $g\cong(0.3-2.0)\times10^{-9}$ |
| $f_{55}$ (V) | 0-10 |
| $v_{311311}$ | $3\;10^{-9}$ |
| Γ (s m J/$C^2$) | $\sim10^{-3}$ |
| $\mu$ ($s^2$ m J/$C^2$) | $8*10^{-14}$ |
| $\rho$ (kg/$m^3$) | 3427 |
| $M$ ($s^2$ J/(m $C^2$)) | $10^{-11}$ |

### A3. Analytical solution for the eigen frequency of the flexoferrons

Following the seminal approach of Tang et al. [67], we consider the case when the mechanical force $N$ and the electric field $E$ in the right side of Eq.(A.3) are Langevin noise fields, which obey the fluctuation-dissipation theorem. Namely, their correlator (averaged over the statistical ensemble in the Fourier space) in the "classical" white noise limit has the form:

$$\langle\tilde{E}(\boldsymbol{k},\omega)\tilde{E}^*(\boldsymbol{k}',\omega')\rangle = 2k_BT\frac{\Gamma}{(2\pi)^2}\delta(\boldsymbol{k}-\boldsymbol{k}')\delta(\omega-\omega'), \tag{A.12a}$$

$$\langle\tilde{N}(\boldsymbol{k},\omega)\tilde{N}^*(\boldsymbol{k}',\omega')\rangle = 2k_BT\frac{\Lambda}{(2\pi)^2}\delta(\boldsymbol{k}-\boldsymbol{k}')\delta(\omega-\omega'). \tag{A.12b}$$

For the "quantum" case $\hbar\omega \geq k_BT$, the correlators are

$$\langle\tilde{E}(\boldsymbol{k},\omega)\tilde{E}^*(\boldsymbol{k}',\omega')\rangle = \frac{\Gamma}{(2\pi)^2}\hbar\omega\coth\left(\frac{\hbar\omega}{2k_BT}\right)\delta(\boldsymbol{k}-\boldsymbol{k}')\delta(\omega-\omega'), \tag{A.12c}$$

$$\langle\tilde{N}(\boldsymbol{k},\omega)\tilde{N}^*(\boldsymbol{k}',\omega')\rangle = \frac{\Lambda}{(2\pi)^2}\hbar\omega\coth\left(\frac{\hbar\omega}{2k_BT}\right)\delta(\boldsymbol{k}-\boldsymbol{k}')\delta(\omega-\omega'). \tag{A.12d}$$

Also $\langle\tilde{E}(\boldsymbol{k},\omega)\rangle = 0$ and $\langle\tilde{N}(\boldsymbol{k},\omega)\rangle = 0$.

Hereinafter we neglect the damping of the elastic phonons, i.e., put $\Lambda = 0$, for the sake of simplicity. For $\Lambda = 0$ the linear polarization response can be found in the first order of the perturbation theory as

$$p_L = \hat{G}E, \;\; \Rightarrow \;\; \tilde{p}_L(\boldsymbol{k},\omega) = \tilde{G}(\boldsymbol{k},\omega)\tilde{E}(\boldsymbol{k},\omega), \tag{A.13a}$$

where $P = P_S + p_L$ and the Green function (propagator) $\hat{G}$ is introduced. The average value

$\langle\tilde{p}_L(\boldsymbol{k},\omega)\rangle = 0$ since $\langle\tilde{E}(\boldsymbol{k},\omega)\rangle = 0$. In accordance with the fluctuation-dissipation theorem and Ornstein-Zernike relation, the Green function spectrum for ferroelectrics is given by the generalized susceptibility, namely $\tilde{G}(\boldsymbol{k},\omega) \cong \tilde{\chi}(\boldsymbol{k},\omega)$ [68]. In the considered case, the dielectric susceptibility spectrum $\tilde{\chi}(\boldsymbol{k},\omega)$ is given by Eq.(A.6b):

$$\tilde{\chi}(\boldsymbol{k},\omega) = \frac{\hat{v}\boldsymbol{k}^4 + \hat{c}\boldsymbol{k}^2 - \rho\omega^2}{\Delta(k,\omega)}, \tag{A.13b}$$

where

$$\Delta(\boldsymbol{k},\omega) = (\alpha_S + \hat{g}\boldsymbol{k}^2 - \mu\omega^2 - i\Gamma\omega)\left(\hat{v}\boldsymbol{k}^4 + \hat{c}\boldsymbol{k}^2 - \rho\omega^2\right) - \left(\hat{f}\boldsymbol{k}^2 - M\omega^2\right)^2 - 4P_S^2(\hat{q}\boldsymbol{k} + 2\hat{z}\boldsymbol{k}P_S^2)^2 \approx$$

$$\approx (\mu\rho - M^2)(\omega_O^2(\boldsymbol{k}) - \omega^2)(\omega_A^2(\boldsymbol{k}) - \omega^2) - i\omega\Gamma\left(\hat{v}\boldsymbol{k}^4 + \hat{c}\boldsymbol{k}^2 - \rho\omega^2\right). \tag{A.13c}$$

The approximate equality in expression (A.13c) follows from Eq.(8).

Equations (A.13) transform in Eq.(7) from Ref.[24] in the limiting case $f_{ij} = 0, M = 0, q_{ij} = z_{ijk} = 0$, namely $\tilde{\chi}(k,\omega) \to \frac{1}{\alpha_S + \hat{g}\boldsymbol{k}^2 - \mu\omega^2 - i\Gamma\omega} \equiv \frac{1}{\mu\left(\omega_p^2(\boldsymbol{k}) - \omega^2\right) - i\Gamma\omega}$, where $\omega_p^2(\boldsymbol{k}) = \frac{1}{\mu}(\alpha_S + \hat{g}\boldsymbol{k}^2)$.

In the second order of perturbation theory the quadratic term appears in the right side of Eq.(A.3b)

$$p = \hat{G}E - [3\beta^* P_S + 10\gamma^* P_S^3 + 21\delta P_S^5]\hat{G}p_L^2 + O(E^3), \tag{A.14}$$

where $P = P_S + p$. Using Fourier transform in Eq.(A.14) we obtain that

$$\tilde{p} = \tilde{\chi}\tilde{E} - [3\beta^* P_S + 10\gamma^* P_S^3 + 21\delta P_S^5]\tilde{\chi}\cdot\widetilde{p_L^2}, \tag{A.15a}$$

The average of Eq.(A.15a) over the statistical ensemble is

$$\langle\tilde{p}(\boldsymbol{k},\omega)\rangle = -[3\beta^* P_S + 10\gamma^* P_S^3 + 21\delta P_S^5]\tilde{\chi}\cdot\langle\widetilde{p_L^2}\rangle, \tag{A.15b}$$

where $\langle\widetilde{p_L^2}(\boldsymbol{k},\omega)\rangle = \int_{-\infty}^{\infty} d\omega' \int_{-\infty}^{\infty} d^3\boldsymbol{k}'\, \tilde{\chi}(\boldsymbol{k}',\omega')\tilde{\chi}^*(\boldsymbol{k}' - \boldsymbol{k}, \omega' - \omega)\langle\tilde{E}(\boldsymbol{k}',\omega')\tilde{E}^*(\boldsymbol{k}' - \boldsymbol{k}, \omega' - \omega)\rangle$ and we used that $\tilde{f}^*(\boldsymbol{k},\omega) \equiv \tilde{f}(-\boldsymbol{k},-\omega)$.

Therefore, the second-order correction in the case of classical white noise is

$$\langle p(\boldsymbol{x},t)\rangle \cong -2k_B T \frac{3\beta^* P_S + 10\gamma^* P_S^3 + 21\delta P_S^5}{\alpha + 3\beta^* P_S^2 + 5\gamma^* P_S^4 + 7\delta P_S^6} \int_{-\infty}^{\infty} \frac{d^3\boldsymbol{k}}{(2\pi)^3} \int_{-\infty}^{\infty} \frac{d\omega}{2\pi} \Gamma|\tilde{\chi}(\boldsymbol{k},\omega)|^2\,, \tag{A.17a}$$

where we used that $\tilde{\chi}(0,0) = \frac{1}{\alpha_S} \equiv \frac{1}{\alpha + 3\beta^* P_S^2 + 5\gamma^* P_S^4 + 7\delta P_S^6}$.

The second-order correction in the case of quantum noise is

$$\langle p(\boldsymbol{x},t)\rangle \cong -\Gamma \frac{3\beta^* P_S + 10\gamma^* P_S^3 + 21\delta P_S^5}{\alpha + 3\beta^* P_S^2 + 5\gamma^* P_S^4 + 7\delta P_S^6} \int_{-\infty}^{\infty} \frac{d\omega}{2\pi} \hbar\omega\coth\left(\frac{\hbar\omega}{2k_B T}\right) \int_{-\infty}^{\infty} \frac{d^3\boldsymbol{k}}{(2\pi)^3} |\tilde{\chi}(\boldsymbol{k},\omega)|^2\,. \tag{A.17b}$$

Since $\Gamma\hbar\omega\coth\left(\frac{\hbar\omega}{2k_B T}\right) \to 2k_B T$ in the classical white noise limit, $\frac{\hbar\omega}{k_B T} \to 0$, Eq.(A.17b) transforms into Eq.(A.17a) in the classical limit. This gives us the opportunity to consider only Eq.(A.17b) in further calculations, which already contains the transformation to the classical limit.

The integration of Eq.(A.17b) in the frequency domain in the case of a very small damping leads to the expression:

$$\langle p\rangle \approx -\Gamma \frac{3\beta^* P_S + 10\gamma^* P_S^3 + 21\delta P_S^5}{\alpha + 3\beta^* P_S^2 + 5\gamma^* P_S^4 + 7\delta P_S^6} \int_{-\infty}^{\infty} \frac{d^3\boldsymbol{k}}{(2\pi)^3} \Xi(\boldsymbol{k})\,, \tag{A.17b}$$

$$\Xi(\boldsymbol{k}) = \int_{-\infty}^{\infty} \frac{d\omega}{2\pi} \xi(\boldsymbol{k},\omega) \hbar\omega \coth\left(\frac{\hbar\omega}{2k_B T}\right) = i \sum_{m=O,A} \mathrm{Res}[\zeta(\omega_m^{\Gamma})], \tag{A.17c}$$

$$\xi(\boldsymbol{k},\omega) = \frac{\left(\hat{\hat{v}}\boldsymbol{k}^4 + \hat{c}\boldsymbol{k}^2 - \rho\omega^2\right)^2}{\left[(\mu\rho - M^2)\left(\omega_O^2(\boldsymbol{k}) - \omega^2\right)\left(\omega_A^2(\boldsymbol{k}) - \omega^2\right)\right]^2 + (\omega\Gamma)^2\left(\hat{\hat{v}}\boldsymbol{k}^4 + \hat{c}\boldsymbol{k}^2 - \rho\omega^2\right)^2}, \tag{A.17d}$$

where $\mathrm{Res}[\zeta(\omega_m^{\mathrm{D}})] = \frac{1}{n!}\lim\left[\frac{d^{n-1}}{d\omega^{n-1}}\{(\omega - \omega_m^{\mathrm{D}})^n \zeta(\omega)\}\right]$ and the integrand $\zeta(\boldsymbol{k},\omega) = \xi(\boldsymbol{k},\omega)\hbar\omega \coth\left(\frac{\hbar\omega}{2k_B T}\right)$ has the $n$-th order pole at the frequency $\omega_m^{\mathrm{D}}$, which has an imaginary part.

The dependences of the function $\xi(k,\omega)$ on the frequency $\omega$ calculated for $\Lambda = 0$, $k_2 = k_3 = 0$, several values of wave vector $k_1$ and damping coefficients $\Gamma$ are shown in **Fig. A1**. Black solid curves are for exact expressions, red dashed and blue dotted curves are for approximations of the integrands near the acoustic and optic frequency values, and magenta dash-dotted curves are the sum of both approximations.

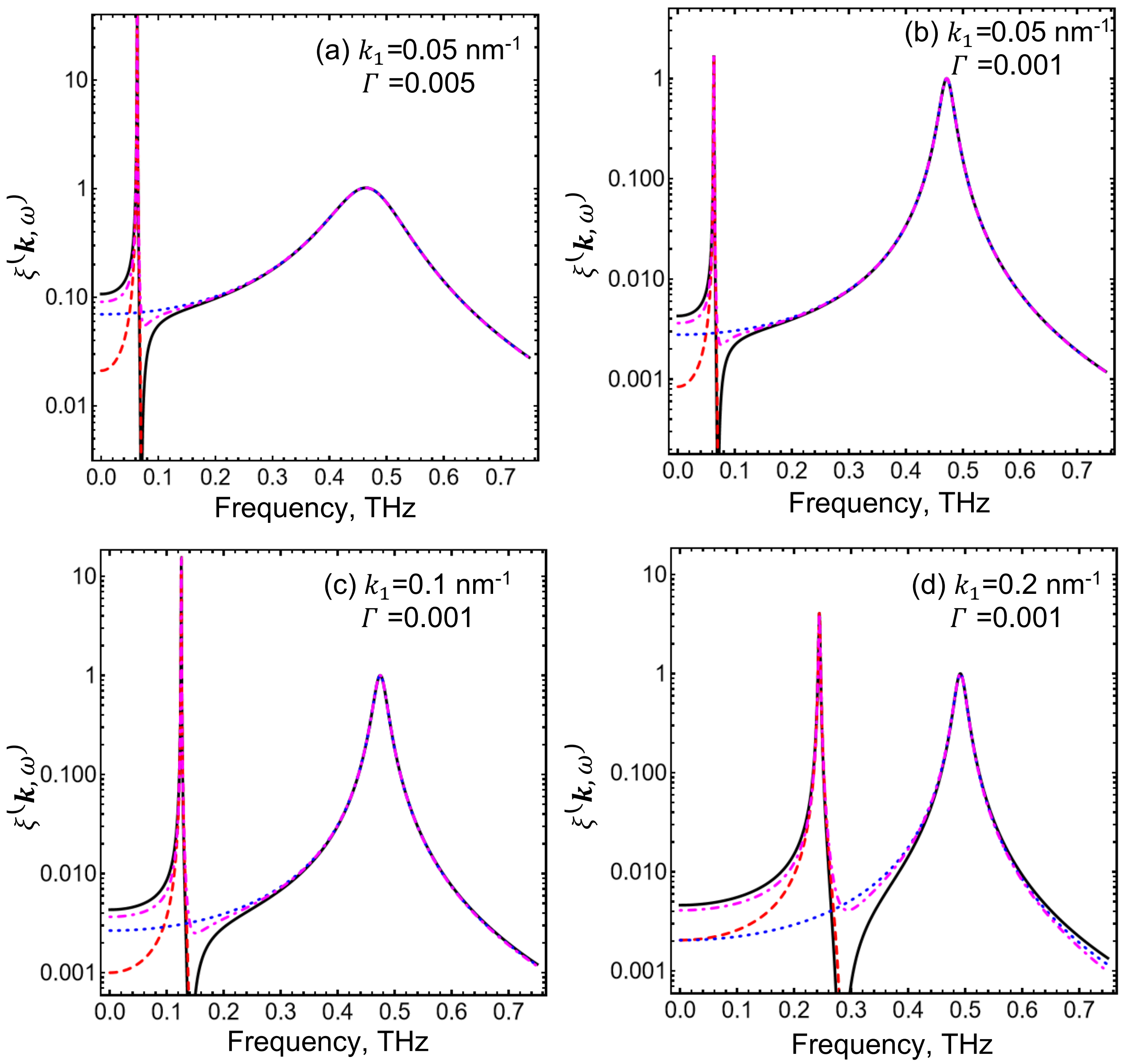

**FIGURE A1.** The dependence of the function $\xi(\boldsymbol{k},\omega)$ on the frequency $\omega$ calculated for $k_2 = k_3 = 0$, several values of wave vector $k_1$ =0.05 nm$^{-1}$ **(a, b)**, 0.1 nm$^{-1}$ **(c)** and 0.2 nm$^{-1}$ **(d)**, damping coefficients $\Gamma$ =0.005 s m J/C$^2$ **(a)** and 0.001 s m J/C$^2$ **(b)-(d)**. Black solid curves are for exact expressions, red dashed and blue dotted curves are for approximations of the integrands near the acoustic and optic frequency values, while magenta dash-dotted curves are the sum of both approximations. The flexoelectric coefficient $f_{55}$ is 1 V, $\Lambda = 0, T = 293$ K. The material parameters of $CuInP_2S_6$ are given in **Table AI**.

In the limiting case $f = 0, M = 0, q = z = 0$ we obtained that $\Xi(k) = \int_{-\infty}^{\infty} \frac{\hbar\omega\coth\left(\frac{\hbar\omega}{2k_BT}\right) d\omega}{\left[\mu(\omega_p^2-\omega^2)\right]^2+(\omega\Gamma)^2} = \int_{-\infty}^{\infty} \frac{\hbar\omega\coth\left(\frac{\hbar\omega}{2k_BT}\right) d\omega}{[\mu(\omega_p^2-\omega^2)+i\Gamma\omega][\mu(\omega_p^2-\omega^2)-i\Gamma\omega]} = \int_{-\infty}^{\infty} \frac{\hbar\omega\coth\left(\frac{\hbar\omega}{2k_BT}\right) d\omega}{\mu^2\prod_{m=1}^{4}(\omega-\omega_m)}$, where $\omega_{1,2} = \frac{i\Gamma}{2\mu} \pm \sqrt{\omega_p^2 - \left(\frac{\Gamma}{2\mu}\right)^2} \approx \frac{i\Gamma}{2\mu} \pm \omega_p$ and $\omega_{3,4} = -\frac{i\Gamma}{2\mu} \pm \sqrt{\omega_p^2 - \left(\frac{\Gamma}{2\mu}\right)^2} \approx -\frac{i\Gamma}{2\mu} \pm \omega_p$, and $\omega_p^2 = \frac{1}{\mu}(\alpha_S + gk^2)$.

Since $\sum_m \mathrm{Res}[f(\omega_m)] = \frac{1}{\mu^2(\omega_1-\omega_2)}\left[\frac{1}{(\omega_1-\omega_3)(\omega_1-\omega_4)} - \frac{1}{(\omega_2-\omega_3)(\omega_2-\omega_4)}\right] \approx \frac{1}{2i\mu\Gamma\omega_p^2}$ . Therefore $\Xi(k) \approx \frac{\hbar}{2\mu\Gamma\omega_p(k)}\coth\left(\frac{\hbar\omega_p}{2k_BT}\right)$ under negligibly small terms proportional to $\left(\frac{\Gamma}{2\mu}\right)^2$, and so Eq.(A.17b) simplifies as

$$\langle p\rangle \approx -\frac{\hbar}{2\mu}\frac{3\beta^*P_S+10\gamma^*P_S^3+21\delta P_S^5}{\alpha+3\beta^*P_S^2+5\gamma^*P_S^4+7\delta P_S^6}\int_{-\infty}^{\infty}\frac{d^3\boldsymbol{k}}{(2\pi)^3}\frac{\coth\left(\frac{\hbar\omega_p(\boldsymbol{k})}{2k_BT}\right)}{\omega_p(\boldsymbol{k})} \xrightarrow[\frac{\hbar\omega}{k_BT}\to 0]{} -k_BT\frac{3\beta^*P_S+10\gamma^*P_S^3+21\delta P_S^5}{\alpha+3\beta^*P_S^2+5\gamma^*P_S^4+7\delta P_S^6}\int_{-\infty}^{\infty}\frac{d^3\boldsymbol{k}}{(2\pi)^3}\frac{1}{\alpha_S+\hat{g}\boldsymbol{k}^2}, \tag{A.18}$$

which is valid in the case of one phonon branch, $\omega_p^2(\boldsymbol{k}) = \frac{1}{\mu}(\alpha_S + \hat{g}\boldsymbol{k}^2)$. Eq.(A.18) is in an agreement with the "classical" limit of the "quantized" spectra Eq.(8) in Ref.[24], because $\coth\left(\frac{\hbar\omega_p}{2k_BT}\right) \approx \frac{2k_BT}{\hbar\omega_p}$ for $\hbar\omega \ll k_BT$ and the Plank constant $\hbar$ canceled in Eq.(A.18) in the classical limit.

We analyze much more complex Eq. (A.17c) considering acoustic and optic phonons. Regarding the very small damping and $\omega_O^2(\boldsymbol{k}) \gg \omega_A^2(\boldsymbol{k})$, we derive approximate expression for $\Xi(\boldsymbol{k})$:

$$\Xi(\boldsymbol{k}) = \int_{-\infty}^{\infty}\frac{\frac{1}{2\pi}\hbar\omega\coth\left(\frac{\hbar\omega}{2k_BT}\right)d\omega}{\left[\frac{(\mu\rho-M^2)(\omega_O^2-\omega^2)(\omega_A^2-\omega^2)}{\rho\omega^2-ck^2-vk^4}\right]^2+(\omega\Gamma)^2} \cong \int_{-\infty}^{\infty}\frac{\frac{1}{2\pi}\left\{\omega^2-\frac{ck^2+vk^4}{\rho}\right\}\left\{\omega_A^2-\frac{ck^2+vk^4}{\rho}\right\}\hbar\omega_A\coth\left(\frac{\hbar\omega_A}{2k_BT}\right)d\omega}{\left[\left(\mu-\frac{M^2}{\rho}\right)(\omega_O^2-\omega_A^2)(\omega_A^2-\omega^2)\right]^2+\left(\left\{\omega_A^2-\frac{ck^2+vk^4}{\rho}\right\}\omega\Gamma\right)^2} +$$

$$\int_{-\infty}^{\infty}\frac{\frac{1}{2\pi}\left\{\omega_O^2-\frac{ck^2+vk^4}{\rho}\right\}^2\hbar\omega_O\coth\left(\frac{\hbar\omega_O}{2k_BT}\right)d\omega}{\left[\left(\mu-\frac{M^2}{\rho}\right)(\omega^2-\omega_O^2)(\omega_O^2-\omega_A^2)\right]^2+\left(\left\{\omega_O^2-\frac{ck^2+vk^4}{\rho}\right\}\omega\Gamma\right)^2} = \left|\begin{array}{l}\int_{-\infty}^{\infty}\frac{d\omega}{\left[\mu(\omega_p^2-\omega^2)\right]^2+(\omega\Gamma)^2} = \frac{\pi}{\mu\Gamma\omega_p^2},\\ \int_{-\infty}^{\infty}\frac{(\omega^2-\omega_0^2)d\omega}{\left[\mu(\omega_p^2-\omega^2)\right]^2+(\omega\Gamma)^2} = \frac{\pi(\omega_p^2-\omega_0^2)}{\mu\Gamma\omega_p^2}\end{array}\right| =$$

$$\frac{\hbar}{2\Gamma\left(\mu-\frac{M^2}{\rho}\right)}\left[\coth\left(\frac{\hbar\omega_A}{2k_BT}\right)\frac{\left|\omega_A^2-\frac{ck^2+vk^4}{\rho}\right|}{(\omega_O^2-\omega_A^2)\omega_A} + \coth\left(\frac{\hbar\omega_O}{2k_BT}\right)\frac{\left|\omega_O^2-\frac{ck^2+vk^4}{\rho}\right|}{(\omega_O^2-\omega_A^2)\omega_O}\right], \tag{A.19}$$

Hereinafter setting $k_2 = k_3 = 0$ and redesignating $k_1 \equiv k$, we also suggest that the integrand consists of two peaks (as shown in **Fig. A1**). Expressions for $\omega^2_{O,A}(k)$ are still given by Eq.(A.8), and $\omega_A(k) \cong \pm\frac{c}{\rho}k$ at very small $k$. Note that approximate expressions in Eq.(A.19) are valid for $\omega^2_A(k) \geq 0$. When $\omega^2_A(k) < 0$, which happens at high values of the flexoelectric coefficient $f_{55} > f_{cr}$, signaling the dynamic instability (possibly transition to the incommensurate phase), other expressions should be used. Hereinafter we consider analytically the case $f_{55} < f_{cr}$.

In the limit of classical white noise $\Xi(\boldsymbol{k}) \to \frac{2k_BT}{2\Gamma(\mu\rho-M^2)}\frac{\hat{v}\boldsymbol{k}^4+\hat{c}\boldsymbol{k}^2}{\omega^2_A\omega^2_O} \equiv \frac{2k_BT}{2\Gamma}\frac{\hat{v}\boldsymbol{k}^4+\hat{c}\boldsymbol{k}^2}{B(k)}$, where $B(\boldsymbol{k}) = \alpha_S\hat{c}\boldsymbol{k}^2 - 4P_S^2(\hat{q}\boldsymbol{k} + 2\hat{z}\boldsymbol{k}P_S^2)^2 + \hat{c}\boldsymbol{k}^2\hat{g}\boldsymbol{k}^2 + \alpha_S\hat{v}\boldsymbol{k}^4 - \left(\hat{f}\boldsymbol{k}^2\right)^2 + \hat{g}\boldsymbol{k}^2\hat{v}\boldsymbol{k}^4$ in accordance with Eq.(A.7c). As expected, Eq.(A.19) simplifies to Eq.(A.18) in the case $f = 0, M = 0, q = z = 0, v = 0$.

Substitution of Eq.(A.19) in Eq.(A.17b) yields the expression for $\langle p\rangle$:

$$\langle p\rangle \approx \int_{-\infty}^{\infty}\frac{d^3\boldsymbol{k}}{(2\pi)^3}\left[\coth\left(\frac{\hbar\omega_A(\boldsymbol{k})}{2k_BT}\right)\delta p_A(\boldsymbol{k}) + \coth\left(\frac{\hbar\omega_O(\boldsymbol{k})}{2k_BT}\right)\delta p_O(\boldsymbol{k})\right], \tag{A.20a}$$

where the “spectral densities” of flexoferrons are

$$\delta p_A(\boldsymbol{k}) \approx \frac{-\hbar}{2(\mu\rho-M^2)}\frac{3\beta^*P_S+10\gamma^*P_S^3+21\delta P_S^5}{\alpha+3\beta^*P_S^2+5\gamma^*P_S^4+7\delta P_S^6}\left|\frac{vk^4+ck^2-\rho\omega_A^2(k)}{\left(\omega_O^2(k)-\omega_A^2(k)\right)\omega_A(k)}\right|, \tag{A.20b}$$

$$\delta p_O(\boldsymbol{k}) \approx \frac{-\hbar}{2(\mu\rho-M^2)}\frac{3\beta^*P_S+10\gamma^*P_S^3+21\delta P_S^5}{\alpha+3\beta^*P_S^2+5\gamma^*P_S^4+7\delta P_S^6}\left|\frac{vk^4+ck^2-\rho\omega_O^2(k)}{\left(\omega_A^2(k)-\omega_O^2(k)\right)\omega_O(k)}\right|. \tag{A.20c}$$

The k-dispersion of spectral densities $\delta p_A(k_1)$ and $\delta p_O(k_1)$ (in $10^{-30}$ C·m) calculated for $k_2 = k_3 = 0$, temperatures 10 K and 293 K, small damping coefficient Γ, and different values of the flexoelectric coefficient $f_{55}$ are shown in **Fig. A2**.

In the limit of classical white noise:

$$\langle p\rangle \approx \frac{-k_BT}{\mu\rho-M^2}\frac{3\beta^*P_S+10\gamma^*P_S^3+21\delta P_S^5}{\alpha+3\beta^*P_S^2+5\gamma^*P_S^4+7\delta P_S^6}\int_{-\infty}^{\infty}\frac{d^3\boldsymbol{k}}{(2\pi)^3}\frac{vk^4+ck^2}{\omega_A^2(k)\omega_O^2(k)} =$$

$$\frac{-k_BT}{\mu\rho-M^2}\frac{3\beta^*P_S+10\gamma^*P_S^3+21\delta P_S^5}{\alpha+3\beta^*P_S^2+5\gamma^*P_S^4+7\delta P_S^6}\int_{-\infty}^{\infty}\frac{d^3\boldsymbol{k}}{(2\pi)^3}\frac{vk^2+c}{\alpha_Sc-4P_S^2\left(q+2zP_S^2\right)^2+(cg+\alpha_Sv-f^2)k^2+gvk^4}. \tag{A.21}$$

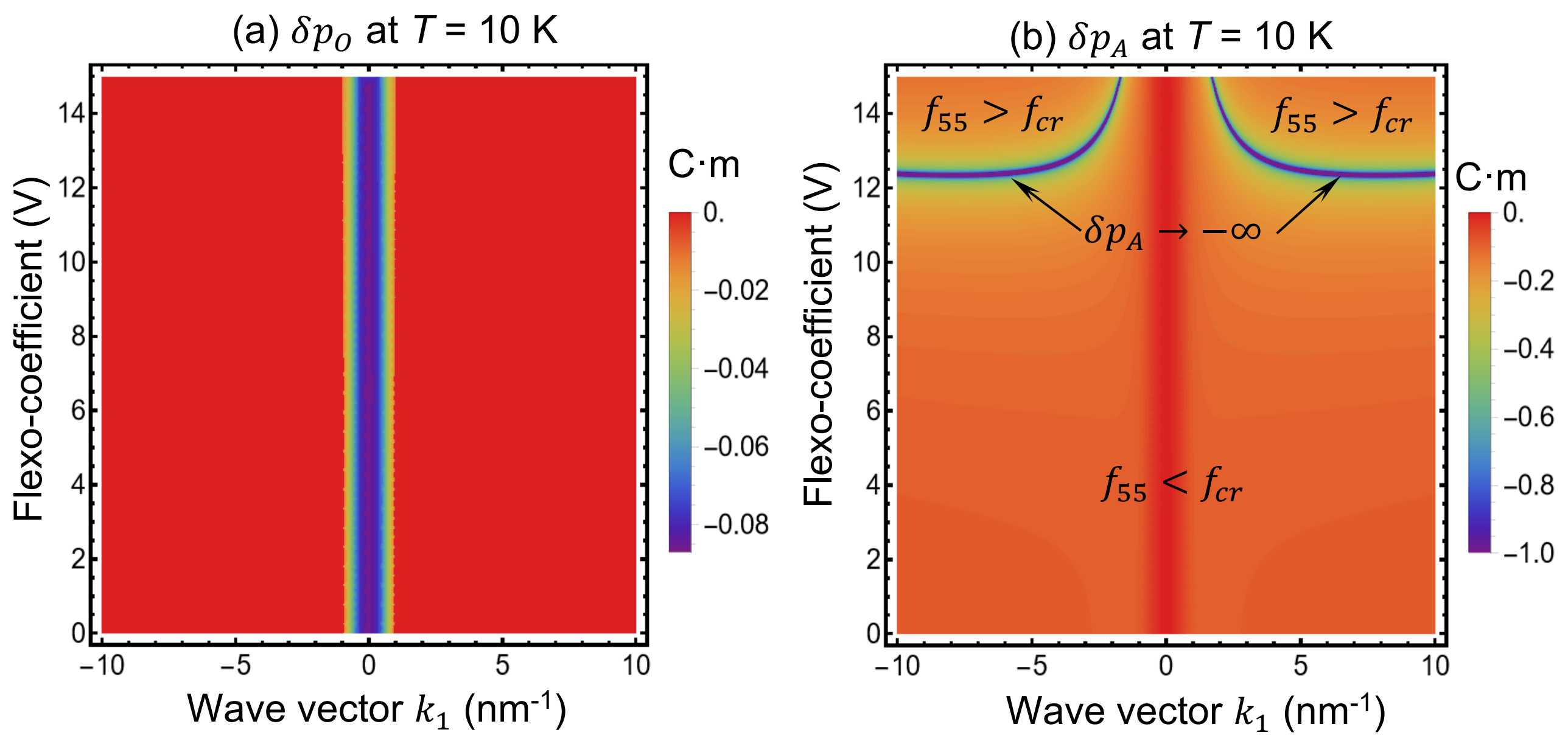

**FIGURE A2.** The color maps of the $\delta p_O(k_1)$ **(a)** and $\delta p_A(k_1)$ **(a)** in dependence on the wavenumber $k_1$ and flexo-coefficient $f_{55}$ calculated for $k_2 = k_3 = 0$, and $T = 10$ K. The material parameters of $CuInP_2S_6$ are listed in **Table AI**.

### A4. The contribution of the flexophonons and flexoferrons to the pyroelectric and electrocaloric response

The temperature dependences of the electrostriction coefficients and LGD coefficients $\beta, \gamma$, and $\delta$, listed in **Table A1**, are significant for temperatures well above the temperature $T_q$. At the same time the contribution of the fluctuations (represented by the flexoferrons) to the pyroelectric response may become significant at the temperatures much lower $T_q$ [24].

The total polarization is $P(T) = P(0) + \Delta P(T)$. Following the approach proposed in Ref.[24], we postulate that the variation, $\Delta P(T)$, and the pyroelectric coefficient, $\Pi(T)$, are

$$\Delta P = \int_{-\infty}^{\infty} \frac{d^3\boldsymbol{k}}{(2\pi)^3}\left(\frac{\delta p_O(k)}{\exp\left(\frac{\hbar\omega_O}{k_B T}\right)-1} + \frac{\delta p_A(k)}{\exp\left(\frac{\hbar\omega_A}{k_B T}\right)-1}\right) \approx \int_{-\infty}^{\infty} \frac{d^3\boldsymbol{k}}{(2\pi)^3}\left(\exp\left(-\frac{\hbar\omega_O}{k_B T}\right)\delta p_O(k) + \exp\left(-\frac{\hbar\omega_A}{k_B T}\right)\delta p_A(k)\right), \tag{A.22a}$$

$$\Pi = -\frac{d}{dT}\Delta P(T) = \int_{-\infty}^{\infty} \frac{d^3\boldsymbol{k}}{(2\pi)^3}\left(\frac{\exp\left(\frac{\hbar\omega_O}{k_B T}\right)\frac{\hbar\omega_O}{k_B T^2}}{\left[\exp\left(\frac{\hbar\omega_O}{k_B T}\right)-1\right]^2}\delta p_O(k) + \frac{\exp\left(\frac{\hbar\omega_A}{k_B T}\right)\frac{\hbar\omega_A}{k_B T^2}}{\left[\exp\left(\frac{\hbar\omega_A}{k_B T}\right)-1\right]^2}\delta p_A(k)\right) \approx$$

$$\frac{\hbar}{k_B T^2}\int_{-\infty}^{\infty} \frac{d^3\boldsymbol{k}}{(2\pi)^3}\left[\exp\left(-\frac{\hbar\omega_O}{k_B T}\right)\omega_O \delta p_O(k) + \exp\left(-\frac{\hbar\omega_A}{k_B T}\right)\omega_A \delta p_A(k)\right], \tag{A.22b}$$

where $\frac{1}{\exp\left(\frac{\hbar\omega}{k_B T}\right)-1}$ is the Boze-Einstein distribution function. Approximate expressions for the integrand in Eqs.(A.22) are valid in the low temperature limit, $\frac{\hbar\omega_A}{k_B T} \gg 1$ and especially $\frac{\hbar\omega_O}{k_B T} \gg 1$.

Temperature dependence of the acoustic and optic phonon ferrons contributions to pyroelectric coefficient Π calculated for $CuInP_2S_6$ parameters, damping coefficient $\Gamma = 0$ and different values of the flexoelectric coefficient $f_{55}$ are shown in **Fig. A3**. It is seen from the figure that the contribution of the optic ferrons becomes much lower (in 2 – 6 orders of magnitude) than the contribution of the acoustic ferrons for the temperatures above (0.5 – 1) K.

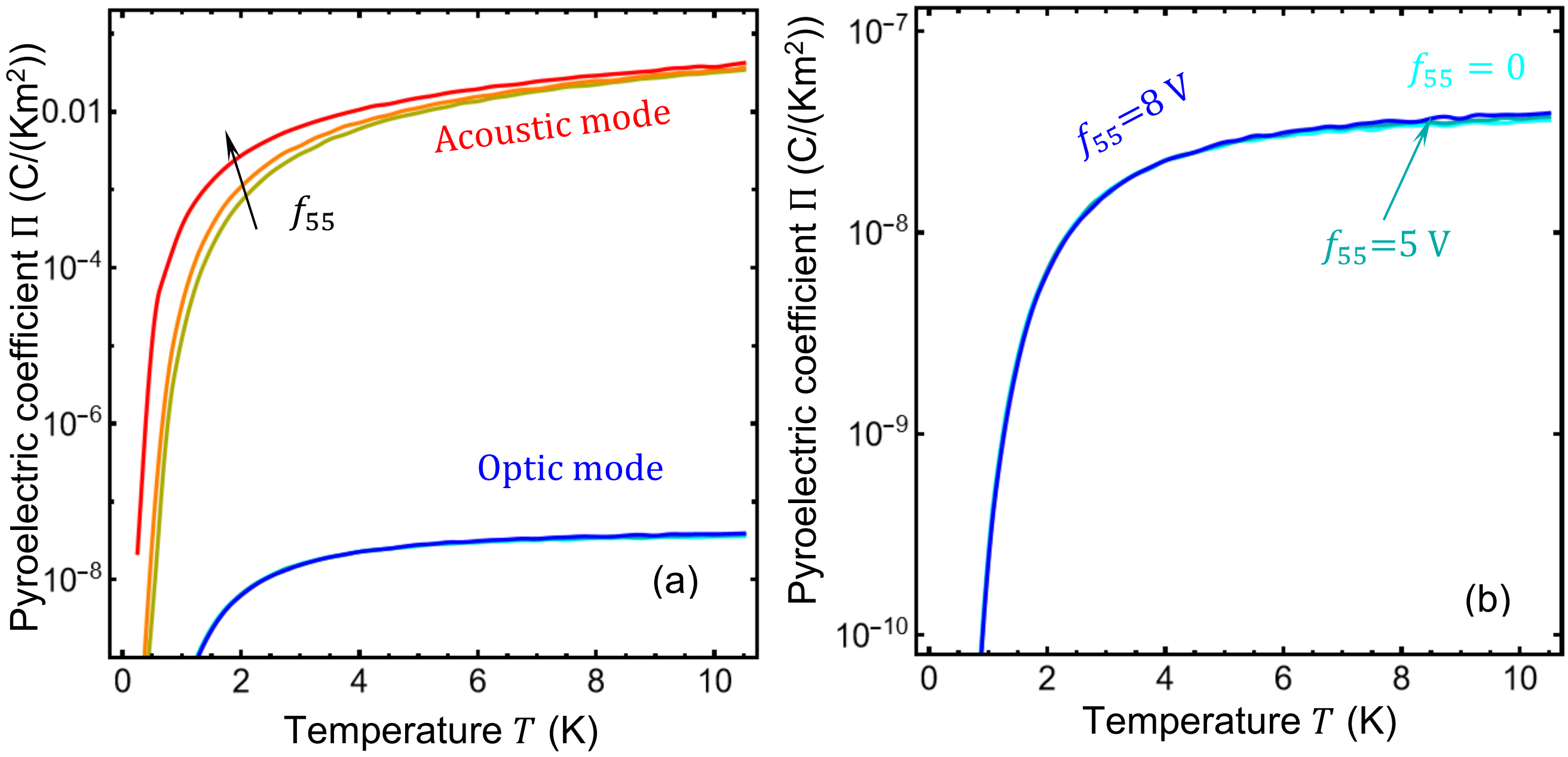

**FIGURE A3.** Temperature dependence of the acoustic **(a)** and optic **(a, b)** ferron contributions to pyroelectric coefficient Π calculated for CIPS material parameters, zero damping and different values of the flexoelectric coefficient $f_{55}$, namely 0, 5V and 8 V, for dark-yellow, orange and red curves (the acoustic mode contribution), blue, dark-cyan and cyan curves (the optic mode contribution). The material parameters of $CuInP_2S_6$ are listed in **Table AI**.

The integrals (A.22) diverge in the 3D **k**-space in the classical limit. In this case one introduces the cross-section $S$ in the $\{x_2, x_3\}$ plane and integrates over $k_1$ only. The product, $\Delta Q = S\Delta P$, can be associated with the pyroelectric charge.

In order to obtain approximate analytical expressions, one can use the method of steepest descent for the integration with exponential functions in Eqs.(A.22). Namely, using that $\exp\left(-\frac{\hbar\omega_O(k)}{k_B T}\right) \approx \exp\left(-\frac{\hbar\omega_O(0)}{k_B T} - \frac{\hbar\omega_O''}{k_B T}\frac{k^2}{2}\right)$ and $\exp\left(-\frac{\hbar\omega_A(k)}{k_B T}\right) \approx \exp\left(-\frac{\hbar|k|}{k_B T}\sqrt{\frac{c}{\rho}}\right)$, we derive the approximate expression from Eq.(A.22a) for the pyroelectric charge

$$\Delta Q(T) \approx \exp\left(-\frac{\hbar\omega_O(0)}{k_B T}\right)\sqrt{\frac{2\pi k_B T}{\hbar\omega_O''(0)}}\,\delta p_O(0) + \frac{k_B T}{\hbar}\sqrt{\frac{\rho}{c}}\,\delta p_A(0) + \mathcal{O}\left[\left(\frac{k_B T}{\hbar}\right)^4\left(\frac{\rho}{c}\right)^2\right]. \quad \text{(A.23a)}$$

Here $\omega_O''(0) = \frac{d^2}{dk^2}\omega_O(k)\Big|_{k\to 0}$.

Since $\delta p_A(0) = 0$, it may seem that only optic flexoferrons contribute to the polarization change at low temperatures. Using that $\delta p_A(k) \approx \frac{-\hbar}{2(\mu\rho - M^2)}\frac{3\beta^* P_S + 10\gamma^* P_S^3 + 21\delta P_S^5}{\alpha + 3\beta^* P_S^2 + 5\gamma^* P_S^4 + 7\delta P_S^6}\frac{v\rho|k|^3}{\omega_O^2(0)}$ and $\delta p_O(0) \approx \frac{-\hbar}{2(\mu\rho - M^2)}\frac{3\beta^* P_S + 10\gamma^* P_S^3 + 21\delta P_S^5}{\alpha + 3\beta^* P_S^2 + 5\gamma^* P_S^4 + 7\delta P_S^6}\frac{\rho}{\omega_O(0)}$, it is possible to estimate both optic and acoustic ferrons contributions to the pyroelectric charge:

$$\Delta Q(T) \approx \frac{-\hbar\rho}{2(\mu\rho - M^2)}\frac{3\beta^* P_S + 10\gamma^* P_S^3 + 21\delta P_S^5}{\alpha + 3\beta^* P_S^2 + 5\gamma^* P_S^4 + 7\delta P_S^6}\left[\frac{1}{\omega_O}\sqrt{\frac{2\pi k_B T}{\hbar\omega_O''}}\exp\left(-\frac{\hbar\omega_O}{k_B T}\right) + \frac{12\,v}{\omega_O^2}\left(\frac{\rho}{c}\right)^2\left(\frac{k_B T}{\hbar}\right)^4\right]. \quad \text{(A.23b)}$$

Hereinafter $\omega_O \equiv \omega_O(0)$ and $\omega_O'' = \frac{d^2}{dk^2}\omega_O(k)\Big|_{k\to 0}$.

Disregarding the temperature dependence of the LGD coefficients, the temperature derivative of the pyroelectric charge, $q(T) = \frac{Q(T)}{dT}$, can be estimated from Eq.(A.23b) as:

$$q(T) \approx \frac{\hbar\rho}{2(\mu\rho - M^2)}\frac{3\beta^* P_S + 10\gamma^* P_S^3 + 21\delta P_S^5}{\alpha + 3\beta^* P_S^2 + 5\gamma^* P_S^4 + 7\delta P_S^6}\left[\left(\frac{1}{\omega_O}\sqrt{\frac{\pi k_B}{2\hbar\omega_O'' T}} + \sqrt{\frac{2\hbar\pi}{k_B\omega_O'' T^3}}\right)\exp\left(-\frac{\hbar\omega_O}{k_B T}\right) + \frac{12\,v}{\omega_O^2}\left(\frac{k_B}{\hbar}\right)^4\left(\frac{\rho}{c}\right)^2 T^3\right]. \quad \text{(A.24)}$$

Expressions (A.23)-(A.24) are valid only at low temperatures, and thus it is more rigorous to calculate the flexoferrons contribution to the pyroelectric response from the initial expression (A.22). If the contribution of the acoustic flexoferrons is neglected and the temperature derivative is taken in Eq.(A.22b), it can lead to the incorrect answer, $q(T) \approx \frac{\hbar\omega_O}{k_B T^2}\Delta Q(T)$. If the contribution of the optic flexoferrons is neglected, $q(T) \approx \frac{1}{T}\Delta Q(T)$.